\documentclass[aps, twocolumn, showpacs, preprintnumbers, superscriptaddress, amsmath, amssymb, prl]{revtex4-1}
\usepackage{amssymb}
\usepackage{mathrsfs}
\usepackage{amsmath}
\usepackage{graphicx}
\usepackage{epstopdf}
\usepackage{float}
\usepackage{multirow}
\usepackage{array}
\usepackage{color}
\usepackage[colorlinks=true, letterpaper=true, pdfstartview=FitV, linkcolor=blue, citecolor=blue, urlcolor=blue]{hyperref}

\makeatletter
\newcommand{\Rmnum}[1]{\expandafter\@slowromancap\romannumeral #1@}
\makeatother
\begin{document}

\title{Strong intervalley correlation induced a magnetic order transition in monolayer $\text{MoS}_{2}$}
\author{Peng Fan}
\affiliation{Theoretical Condensed Matter Physics and Computational Materials Physics Laboratory, College of Physical Sciences, University of Chinese Academy of Sciences, Beijing 100049, China}
\affiliation{School of Electronic, Electrical and Communication Engineering, University of Chinese Academy of Sciences, Beijing 100049, China}

\author{Zhen-Gang Zhu}
\email{zgzhu@ucas.ac.cn}
\affiliation{School of Electronic, Electrical and Communication Engineering, University of Chinese Academy of Sciences, Beijing 100049, China}
\affiliation{Theoretical Condensed Matter Physics and Computational Materials Physics Laboratory, College of Physical Sciences, University of Chinese Academy of Sciences, Beijing 100049, China}
\affiliation{CAS Center for Excellence in Topological Quantum Computation, University of Chinese Academy of Sciences, Beijing 100190, China}

\begin{abstract}
  In this work, we study a model for monolayer molybdenum disulfide with including the intravalley and intervalley electron-electron interaction.  We solve the model at a self-consistent mean-field level and get three solutions $L_{0}$, $L_{+}$ and $L_{-}$. As for $L_{0}$, the spin polarizations are opposite at $\textbf{K}$ and $\textbf{K}^{\prime}$ valley and the total magnetization is zero. $L_{\pm}$ describe two degenerate spin-polarized states, and the directions of polarization are opposite for the states of $L_{+}$ and $L_{-}$. Based on these results, the ground state can be deduced to be spin polarized in domains in which their particular states can be randomly described by $L_{+}$ or $L_{-}$. Therefore, a zero net magnetization is induced for zero external magnetic field $\mathbf{B}$, but a global ferromagnetic ground state for a nonzero $\mathbf{B}$. We estimate the size of domains as several nanometers. As the increase of the chemical potential, the ground state changes between $L_{0}$ and $L_{\pm}$, indicating first order phase transitions at the borders, which is coincident with the observation of photoluminescence experiments in the absence of the external magnetic field [J. G. Roch \emph{et al.}, Phys. Rev. Lett. {\bf 124}, 187602 (2020)].
\end{abstract}

\maketitle
Transition-metal dichalcogenides (TMDCs)~\cite{Manzeli2017,Mak2016} are a class of materials of the type $\text{MX}_{2}$, where $\text{M}$ is a transition-metal atom ($\text{Mo}$, $\text{W}$, $\text{V}$, $\text{Hf}$, etc) and $\text{X}$ is a chalcogen atom (S, Se, Te, etc). In recent decades, interest is grown rapidly in TMDCs due to their impressive electronic~\cite{Manzeli2017,Wang2012, Mak2010, Wilson1969, Wilson1974, Ugeda2016}, optical~\cite{Mak2016, Wang2012} and mechanical properties~\cite{Bertolazzi2011}, and the broad application to electronics~\cite{Radisavljevic2011,Das2014,Marega2020}, spintronics~\cite{Zibouche2014,Xiao2012}, valleytronics~\cite{Zeng2012,Schaibley2016}, optoelectronics~\cite{Mak2016,Koppens2014}, and sensing~\cite{Wang2020}. When bulk TMDCs are thinned to monolayers, correlation effects become much more important than that in the bulk, because the three dimension Coulomb interaction is only screened in two dimensions, which results in a weak dielectric screening~\cite{Chernikov2014}. Many experiments have demonstrated the existence of strong electron-electron (e-e) interaction in monolayer TMDCs (ML-TMDCs), including interaction induced giant paramagnetic response in ML-$\text{MoSe}_{2}$~\cite{Back2017}, new photoluminescence peaks in ML-$\text{WX}_{2}$ (X $=$ S, Se)~\cite{Shang2015,You2015}, enhanced valley magnetic response and quantum Hall states sequence transition in ML-$\text{WSe}_{2}$~\cite{Wang2018,Movva2017}. Optical susceptibility measurements of the molybdenum disulfide ($\text{MoS}_{2}$) monolayer in van der Waals heterostructure provided by Roch \emph{et al}. show that e-e interactions, especially the intervalley exchange interaction, result in a first order phase transition from a spin unpolarized ground state to a spin polarized state in presence of an external magnetic field $\mathbf{B}$~\cite{Roch2019,Roch2020,Dery2016}. In the photoluminescence spectrum, an abrupt change marks this first order phase transition when the trion peak ($\text{X}^{-}$) evolves into the Mahan exciton peak ($\text{Q}$)~\cite{Roch2020}. This first order phase transition attributes to the nonanalytic correction in the free energy~\cite{Roch2020,Miserev2019}. Without $\mathbf{B}$, the same abrupt change is still observed, which implies that a magnetic order transition occurs like the case of nonzero $\mathbf{B}$~\cite{Roch2020}. However, the total magnetization is zero in the whole process, which seems to indicate that the transition of the magnetic order doesn't occur. It is confusing. Roch \emph{et al}.~\cite{Roch2020} proposed that the fluctuation between ``puddles" of the spin up and spin down leads to the zero total magnetization at low electron density. However, there is no theoretical demonstration of the ``puddles" (the degenerate spin polarized states). In previous theoretical studies~\cite{Donck2018,Braz2018} intervalley e-e interaction was ignored and the spin-spin couplings in the intravalley and intervalley were not appreciated, which play a vital role, as shown by our results, in determining the properties of the ground state. We are motivated by the zero magnetic field experimental observations and the lacking of the theoretical explanation. Therefore, we focus on this case and try to understand the peculiar observations in experiments. In this paper, we study a model for ML-$\text{MoS}_{2}$ with including the intravalley and intervalley Coulomb interaction, based on the low-energy noninteracting Hamiltonian derived in previous studies~\cite{Xiao2012} and develop a self-consistent mean field method (SCMFM), emphasizing the effective intervalley spin-spin couplings. It is found that the ground state is composed of two degenerated spin-polarized states at a certain electron density, giving rise to a zero total magnetization. By tuning the electron density via the chemical potential, a first order phase transition occurs between the unpolarized state to the spin-polarized states, which is consistent with the experiment~\cite{Roch2020}.

\begin{figure}[hpt!]
	\begin{center}
		\includegraphics[width=1.0\columnwidth]{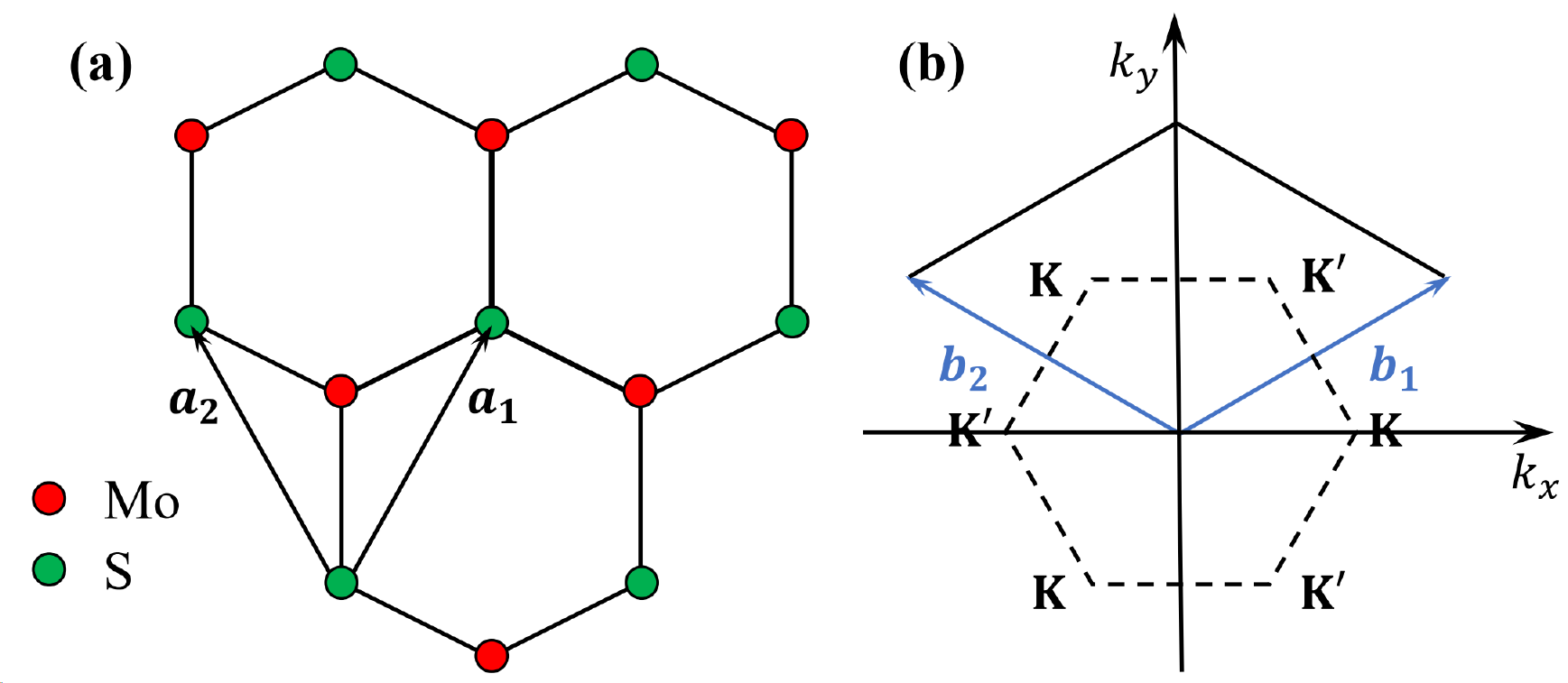}
	\end{center}
	\caption{(Color online) (a) Honeycomb lattice of the monolayer $\text{MoS}_{2}$. The red dot is Mo and the green dot is S. $\bm{a}_{1}$ and $\bm{a}_{2}$ are the primitive vectors. (b) Brillouin zone (BZ) of the honeycomb lattice. $\bm{b}_{1}$ and $\bm{b}_{2}$ are the primitive vectors of the reciprocal lattice. }   \label{Fig1}
\end{figure}

Fig. \ref{Fig1} shows the crystal structure of ML-$\text{MoS}_{2}$ and its first Brillouin zone (BZ)~\cite{Kadantsev2012}. The minima of the conduction band are located at the corners ($\textbf{K}, \textbf{K}^{\prime}$). For the description of the
noninteracting case, we use the effective Hamiltonian of ML-$\text{MoS}_{2}$ around the Dirac cones ~\cite{Xiao2012,Note} as
\begin{equation} \label{1}
\hat{\text{H}}_{0} = at\left(\tau k_{x} \sigma_{x} + k_{y} \sigma_{y}\right) + \frac{ \Delta}{2}\sigma_{z} - \lambda\tau\frac{\sigma_{z} - 1}{2} \hat{s}_{z},
\end{equation}
where $\tau = \pm 1$ are valley indexes for $\textbf{K}$ and $\textbf{K}^{\prime}$ (see Fig.~\ref{Fig1}). The spin splitting caused by spin orbit coupling is $2\lambda$. $\sigma_{\alpha}$ ($\alpha = x, y, z$) are the Pauli matrices. $a$ is the lattice constant. $t$ is the hopping integral. $\Delta$ is the energy gap between the conduction band and valence band (when $\lambda=0$). $\hat{s}_{z}$ is the $z$ component of the spin operator. For convenience, BZ is chosen as the diamond region in the following calculation (see Fig.~\ref{Fig1} (b)). The energy eigenvalues of the Hamiltonian are~\cite{Yu2015}
\begin{equation} \label{energy_spectrum}
E_{n \tau s} = \lambda\tau s/2 \pm \sqrt{(atk)^{2} + \left[(\Delta - \lambda\tau s)/2 \right]^{2}},
\end{equation}
where $s =\pm 1$ are the spin indexes for spin up and down respectively. The up plus (bottom minus) sign in Eq. (\ref{energy_spectrum}) denotes the conduction (valence) band [c (v)]. $k$ is the module of the wave vector. The corresponding eigenstates are denoted as $| n \tau \bm{k} s \rangle$, where $n = \text{c}$ or $\text{v}$.

The Coulomb interaction between electrons is
\begin{equation}
\text{V}(\bm{r}_{1} - \bm{r}_{2}) = \frac{e^2}{4\pi\epsilon_{0}} \frac{1}{|\bm{r}_{1} - \bm{r}_{2}|},
\end{equation}
where $e$ is the elementary charge, $\epsilon_{0}$ is the vacuum permittivity. It is secondly quantized in the $| n \tau \bm{k} s \rangle$ representation~\cite{Bruus2004}
\begin{equation} \label{v_definition}
\hat{\text{V}} = \frac{1}{N} \sum_{s_{1}s_{2}} \sum_{\substack{\bm{k}_{1}\bm{k}_{2}\\ \bm{k}_{3} \bm{k}_{4}}} \sum_{\substack{n_{1}n_{2} \\ n_{3} n_{4}}}
\sum_{\substack{\tau_{1}\tau_{2} \\ \tau_{3} \tau_{4}}} \text{V}_{\text{int}} a^{n_{1}\tau_{1}\dagger}_{\bm{k}_{1} s_{1}} a^{n_{2}\tau_{2}\dagger}_{\bm{k}_{2} s_{2}}
a^{n_{3}\tau_{3}}_{\bm{k}_{3} s_{2}}
a^{n_{4}\tau_{4}}_{\bm{k}_{4} s_{1}},
\end{equation}
where $\text{V}_{\text{int}}$ denotes the strength of the e-e interaction and $N$ is the number of unit cells. $a^{n\tau\dagger}_{\bm{k} s}$ $(a^{n \tau}_{\bm{k} s})$ is the creation (annihilation) operator in $| n \tau \bm{k} s \rangle$ state. There are three kinds of e-e interaction: interaction between the conduction electrons, interaction between the valence electrons and the interaction between the conduction electrons and the valence electrons. Here, we only take the interaction between the conduction electrons into consideration and eliminate the letter c, which is used to mark the conduction band, in the following formulas for convenience. The strength of the e-e interaction in the conduction band is written as
\begin{equation}
\text{V}_{\text{int}} = \frac{1}{2N} \langle  \tau_{1} \bm{k}_{1} s_{1}, \tau_{2} \bm{k}_{2} s_{2}| \text{V} (\bm{r}_{1} - \bm{r}_{2}) |  \tau_{4} \bm{k}_{4} s_{1},  \tau_{3} \bm{k}_{3} s_{2} \rangle.
\end{equation}
$\hat{\text{V}}$ in Eq. (\ref{v_definition}) is thus expanded explicitly and approximated as
\begin{equation} \label{v_approx}
\hat{\text{V}} \approx \hat{\text{V}}_{\text{intra}} + \hat{\text{V}}_{\text{inter}},
\end{equation}
where
\begin{align}
\hat{\text{V}}_{\text{intra}} &=
\frac{1}{N}
\sum_{\bm{k}_{1} \bm{k}_{2}}
\sum_{\tau s}
U
a^{ \tau \dagger}_{\bm{k}_{1} s}
a^{ \tau \dagger}_{\bm{k}_{2} \bar{s}}
a^{ \tau }_{\bm{k}_{2} \bar{s}}
a^{ \tau }_{\bm{k}_{1} s}, \\
\hat{\text{V}}_{\text{inter}} &=
\frac{1}{N}
\sum_{\bm{k}_{1}\bm{k}_{2}}
\sum_{\tau s_{1} s_{2}}
U^{\prime}
a^{ \tau \dagger}_{\bm{k}_{1} s_{1}}
a^{ \bar{\tau} \dagger}_{\bar{\bm{k}}_{2} s_{2}}
a^{ \tau  }_{\bm{k}_{1} s_{2}}
a^{ \bar{\tau} }_{\bar{\bm{k}}_{2} s_{1}}.
\end{align}
$\hat{\text{V}}_{\text{intra}}$ and $\hat{\text{V}}_{\text{inter}}$ denote the intravalley and intervalley e-e interaction respectively. $U$ and $U^{\prime}$ are the  strengths of the corresponding e-e interaction,
\begin{align}
U &= \frac{1}{2 N} \langle \tau \bm{k}_{1},\tau \bm{k}_{2} | \text{V} \left( \bm{r}_{1} - \bm{r}_{2} \right) | \tau \bm{k}_{1},\tau \bm{k}_{2} \rangle, \\
U^{\prime} &= \frac{1}{2 N} \langle \tau \bm{k}_{1},\bar{\tau} \bar{\bm{k}}_{2} | \text{V} \left( \bm{r}_{1} - \bm{r}_{2} \right) | \bar{\tau} \bar{\bm{k}}_{2},\tau \bm{k}_{1} \rangle.
\end{align}
$\bar{\tau}$ $(\bar{s})$ represents the opposite valley (spin) of $\tau$ $(s)$. $\bm{k}$ ($\bar{\bm{k}}$) indicates the relative wave vector with respect to the minimum of $\tau$ ($\bar{\tau}$) valley. Quantitatively, it has been estimated in the static screening limit that due to the small Bohr radius the intervalley e-e interaction is comparable to the intravalley interaction even at high electron density~\cite{Dery2016}. Hence, it is necessary to take the intervalley e-e interaction into consideration when one deals with the Coulomb interaction in TMDCs~\cite{Dery2016}.
For the purpose of a qualitative discussion, $U$ and $U^{\prime}$ are regarded as constants. Details of the above approximation are shown in appendix A.

\begin{figure}[t!]
	\begin{center}
		\includegraphics[width=1.0\columnwidth]{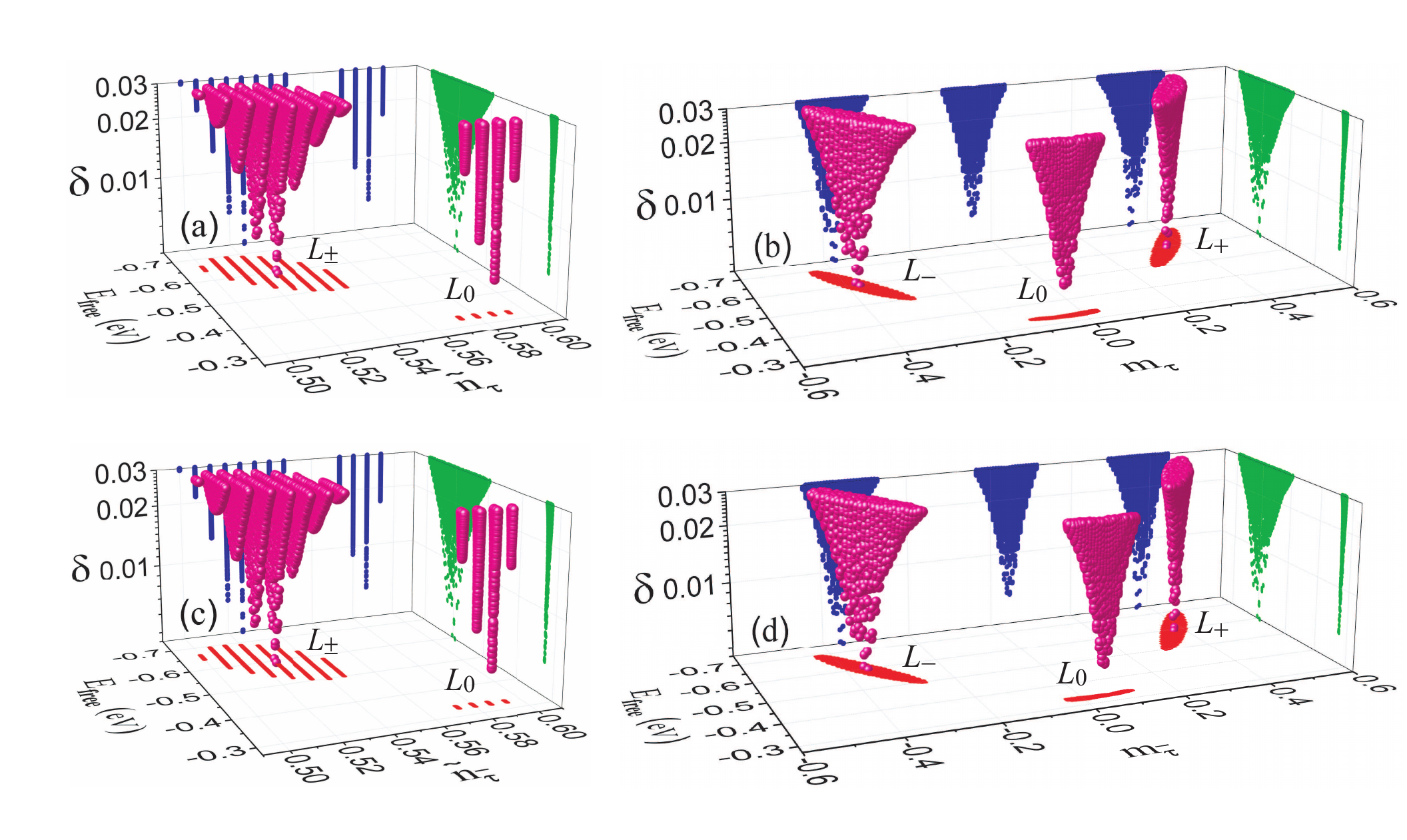}
	\end{center}
	\caption{(Color online) Solutions of MFEs. The parameter space is ergodic by griding the space $150\times150\times150\times150$. We omit the points which have a derivation $\delta > 0.03$. The derivation $\delta$ of $\tilde{n}_{\tau}$, $m_{\tau}$, $\tilde{n}_{\bar{\tau}}$, and $m_{\bar{\tau}}$, are shown in (a), (b), (c), and (d). $L_{0}$, $L_{+}$ and $L_{-}$ are the three solutions. The points in plane are projections of the red spheres. we take $U = 1.0$ eV, $U^{\prime}=0.4$ eV, and $\mu=1.9$ eV.} \label{solutions_of_MFEs}
\end{figure}

We apply the mean field approximation~\cite{Bruus2004} to $\hat{\text{V}}_{\text{intra}}$ and $\hat{\text{V}}_{\text{inter}}$ respectively. As for $\hat{\text{V}}_{\text{intra}}$, it reads
\begin{equation}
\hat{\text{V}}^{\text{MF}}_{\text{intra}}
\approx
\sum_{\bm{k} s}
\left(
\mathbb{U}_{\tau s}
a^{\tau \dagger}_{\bm{k} s}
a^{\tau }_{\bm{k} s}
+
\mathbb{U}_{\bar{\tau} s}
a^{\bar{\tau} \dagger}_{\bar{\bm{k}} s}
a^{\bar{\tau} }_{\bar{\bm{k}} s}
\right),
\end{equation}
where
\begin{equation} \label{MF_1}
\mathbb{U}_{\tau s} =
\frac{2}{N}
\sum_{\bm{k}}
U
\langle
n^{\tau }_{\bm{k} \bar{s}}
\rangle
,\,\,\,\,\
\mathbb{U}_{\bar{\tau} s} =
\frac{2}{N}
\sum_{\bm{k}}
U
\langle
n^{\bar{\tau} }_{\bar{\bm{k}} \bar{s}}
\rangle.
\end{equation}
$n^{\tau}_{\bm{k} s} = a^{\tau \dagger}_{\bm{k} s}
a^{\tau}_{\bm{k} s}$ is the particle number operator. In this paper, we merely consider the zero temperature case. Therefore, $\langle\cdots\rangle$ means the ground state average. In terms of the spin operators, $
S^{z}_{\tau \bm{k}} = \frac{1}{2} ( n^{\tau}_{\bm{k} \uparrow} - n^{\tau}_{\bm{k} \downarrow}) $
,
$S^{+}_{\tau \bm{k}} = a^{\tau \dagger}_{\bm{k} \uparrow} a^{\tau}_{\bm{k} \downarrow}$
and
$S^{-}_{\tau \bm{k}} = a^{\tau \dagger}_{\bm{k} \downarrow} a^{\tau}_{\bm{k} \uparrow}$,
$\hat{\text{V}}_{\text{inter}}$ is rewritten as
\begin{equation} \label{v_inter}
\hat{\text{V}}_{\text{inter}} =
-\frac{U^{\prime}}{N}
\sum_{\bm{k}_{1}\bm{k}_{2}}
\left(
n^{\tau}_{\bm{k}_{1}} n^{\bar{\tau}}_{\bar{\bm{k}}_{2}} +
4\bm{S}_{\tau \bm{k}_{1}} \cdot \bm{S}_{\bar{\tau} \bar{\bm{k}}_{2}}  \right),
\end{equation}
where $n^{\tau}_{\bm{k}} = \sum_{s} n^{\tau}_{\bm{k} s}$. $\bm{S}$ is the spin operator and
\begin{equation}
\bm{S}_{\tau \bm{k}_{1}} \cdot \bm{S}_{\bar{\tau} \bar{\bm{k}}_{2}}
=
S^{z}_{\tau \bm{k}_{1}} S^{z}_{\bar{\tau} \bar{\bm{k}}_{2}} + \frac{1}{2} \left( S^{+}_{\tau \bm{k}_{1}} S^{-}_{\bar{\tau} \bar{\bm{k}}_{2}} + S^{-}_{\tau\bm{k}_{1}} S^{+}_{\bar{\tau} \bar{\bm{k}}_{2}} \right).
\end{equation}
In Eq. (\ref{v_inter}), the first and second term give the intervalley density-density interaction and the intervalley spin-spin coupling. We apply mean field approximation to Eq. (\ref{v_inter}) and obtain
\begin{equation} \label{v_inter_mf}
\hat{\text{V}}^{\text{MF}}_{\text{inter}}
\approx
- \sum_{\bm{k}}
\left(
\mathbb{U}^{\prime}_{\bar{\tau}}
n^{\bar{\tau}}_{\bar{\bm{k}}}
+
\mathbb{U}^{\prime}_{\tau}
n^{\tau}_{\bm{k}}
+
\mathbb{M}_{\bar{\tau}}
S^{z}_{\bar{\tau} \bar{\bm{k}}}
+
\mathbb{M}_{\tau}
S^{z}_{\tau \bm{k}}
\right),
\end{equation}
where
\begin{equation}
\mathbb{U}^{\prime}_{\bar{\tau}(\tau)} =
\frac{1}{N}
\sum_{\mathbf{k}(\bar{\mathbf{k}})}
U^{\prime}
\langle
n^{\tau(\bar{\tau})}_{\mathbf{k}(\bar{\mathbf{k}})}
\rangle,\,\,\,
\mathbb{M}_{\bar{\tau}(\tau)} =
\frac{4}{N}
\sum_{\mathbf{k}(\bar{\mathbf{k}})}
U^{\prime}
\langle
S^{z}_{\tau \mathbf{k}(\bar{\tau}\bar{\mathbf{k}})}
\rangle.
\notag
\end{equation}
In Eq. (\ref{v_inter}), the direction of $\bm{S}$ is chosen as the $z$-axis. By defining
\begin{equation} \label{MF_4}
\mathbb{X}_{\tau s} = \mathbb{U}^{\prime}_{\tau} + \frac{1}{2} s\mathbb{M}_{\tau}
,\,\,\,\,\
\mathbb{X}_{\bar{\tau} s} = \mathbb{U}^{\prime}_{\bar{\tau}} + \frac{1}{2}s\mathbb{M}_{\bar{\tau}},
\end{equation}
where $s=\pm1$ for states with spins parallel and antiparallel to the $\bm{S}$, and we obtain  
\begin{equation}
\hat{\text{V}}^{\text{MF}}_{\text{inter}} \approx
-\sum_{\bm{k}s}
\left(
\mathbb{X}_{\tau s}
a^{\tau\dagger}_{\bm{k} s}
a^{\tau}_{\bm{k} s}
+
\mathbb{X}_{\bar{\tau} s}
a^{\bar{\tau}\dagger}_{\bar{\bm{k}} s} a^{\bar{\tau}}_{\bar{\bm{k}} s}\right).
\end{equation}
Therefore, the mean field approximation of the interaction, i.e. $\hat{\text{V}}_{\text{MF}}=\hat{\text{V}}^{\text{MF}}_{\text{intra}}+\hat{\text{V}}^{\text{MF}}_{\text{inter}}$, is
\begin{equation}
\hat{\text{V}}_{\text{MF}} =
\sum_{\bm{k} s}
\left(
\mathbb{F}_{\tau s}
a^{\tau\dagger}_{\bm{k} s}
a^{\tau}_{\bm{k} s}
+
\mathbb{F}_{\bar{\tau} s}
a^{\bar{\tau} \dagger}_{\bar{\bm{k}} s}
a^{\bar{\tau}}_{\bar{\bm{k}} s}
\right).
\end{equation}
$\mathbb{F}_{\tau s}$ and $\mathbb{F}_{\bar{\tau} s}$ are the effective mean fields, which read
\begin{equation} \label{MF_5}
\mathbb{F}_{\tau s} =
\mathbb{U}_{\tau s}
-
\mathbb{X}_{\tau s}
,\,\,\,\,\,
\mathbb{F}_{\bar{\tau} s} =
\mathbb{U}_{\bar{\tau} s}
-
\mathbb{X}_{\bar{\tau} s}
.
\end{equation}
The total mean field Hamiltonian reads
\begin{equation}
\text{H}_{\text{total}}^{\text{MF}} = \sum_{\bm{k} s}
\left(
\mathbb{E}_{\text{c}\tau s} \left(\bm{k}\right)
a^{\tau\dagger}_{\bm{k}s}
a^{\tau}_{\bm{k}s}
+
\mathbb{E}_{\text{c}\bar{\tau}s} \left(\bar{\bm{k}}\right)
a^{\bar{\tau} \dagger}_{\bar{\bm{k}} s}
a^{\bar{\tau}}_{\bar{\bm{k}} s}
\right),
\end{equation}
where the energy spectrum
\begin{equation}
\mathbb{E}_{\text{c}\tau (\bar{\tau}) s} \left(\bm{k}\right) =
E_{\text{c} \tau (\bar{\tau})s} \left(\bm{k} \right)
+ \mathbb{F}_{\tau (\bar{\tau})s} - \mu,
\label{MF_6}
\end{equation}
where $\mu$ is the chemical potential. In above mean field approximation we omit the constant terms, which do not affect our general discussions and qualitative conclusions. The constant terms neglected in the calculations merely shift all energy bands simultaneously. This leads us a zero-energy redefinition. This shift cannot affect the determination of the solutions which are determined by the parameters that are not entangled with the absolute energies but the relative energy with respect to the zero-energy. It is easy then to calculate the free energy
\begin{equation} \label{free energy}
\text{E}_{\text{free}} = \sum_{\tau s} \int \mathbb{E}_{\text{c}\tau s}\left(\bm{k}\right) d\bm{k}.
\end{equation}
The detailed calculations of $\text{E}_{\text{free}}$ can be found in appendix C. In order to calculate the effective mean fields, averages $\langle n^{\tau}_{\bm{k} s}\rangle$ needs to be calculated. We thus introduce
\begin{equation} \label{occ_ratio}
\tilde{n}^{\tau}_{ s } = \frac{1}{N} \sum_{\bm{k}} \langle n^{\tau}_{\bm{k} s}\rangle.
\end{equation}
The total electron number per unit cell at $\tau$ valley is $\tilde{n}_{\tau} = \sum_{s} \tilde{n}^{\tau}_{ s }$, lying in a domain of $[0,1]$. It is convenient for the following discussion to define the valley magnetization as
\begin{equation} \label{m_defination}
m_{\tau} = \tilde{n}^{\tau}_{\uparrow} - \tilde{n}^{\tau}_{\downarrow}
,\,\,\,\,\,
m_{\bar{\tau}} = \tilde{n}^{\bar{\tau}}_{\uparrow} - \tilde{n}^{\bar{\tau}}_{\downarrow},
\end{equation}
which indicate the valley spin polarization. The total magnetization is then $m = m_{\tau} + m_{\bar{\tau}}$.
When the ground state is spin polarized, the total magnetization $m \neq 0$. In contrast, $m = 0$. In terms of $\tilde{n}_{\tau}$ and $m_{\tau}$, the mean field is rewritten as
\begin{equation} \label{MF_total}
\mathbb{F}_{\tau s} = U \left( \tilde{n}_{\tau} + \bar{s} m_{\tau} \right) - \text{U}^{\prime} \left( \tilde{n}_{\bar{\tau}} + s m_{\bar{\tau}} \right).
\end{equation}
The gap of the spin splitting of the conduction band is readily obtained
\begin{equation} \label{egap}
\Delta^{\text{c}\tau}_{E}\left( \textbf{k} \right)
=
E_{\text{c}\tau\uparrow}\left(\bm{k}\right)
-
E_{\text{c}\tau\downarrow}\left(\bm{k}\right) + \Delta^{\tau}_{\mathbb{F}},
\end{equation}
where $\Delta^{\tau}_{\mathbb{F}} = -2 U m_{\tau}  - 2 U^{\prime} m_{\bar{\tau}}$, which shows the influence of the e-e interaction on the spin splitting of the conduction band, and indicates the renormalization of the conduction band minimum (CBM). The renormalized position of CBM is self-consistently calculated. Parameters $\tilde{n}_{\tau}$, $m_{\tau}$, $\tilde{n}_{\bar{\tau}}$,  and $m_{\bar{\tau}}$ constitute a four dimension parameter space. Any point in the space is denoted as a vector $\left(\tilde{n}_{\tau},  m_{\tau}, \tilde{n}_{\bar{\tau}}, m_{\bar{\tau}}\right)$. At this stage, we have obtained all of mean field equations, which are solved numerically and self-consitently. The procedure of the numerical calculation is as following. At first, we give a set of values for $\tilde{n}_{\tau}$, $m_{\tau}$, $\tilde{n}_{\bar{\tau}}$, and $m_{\bar{\tau}}$, which corresponds to a vector $\textbf{P}_{\text{given}}$ in the parameter space.
Then, the effective mean field $\mathbb{F}_{\tau s}$ is obtained by substituting $\tilde{n}_{\tau}$, $m_{\tau}$, $\tilde{n}_{\bar{\tau}}$, and $m_{\bar{\tau}}$ into Eq. (\ref{MF_total}). Utilizing Eq. (\ref{MF_6}), we get the energy spectrum. Finally, we calculate $\text{E}_{\text{free}}$ by Eq. (\ref{free energy}) and update $\tilde{n}_{\tau}$,  $m_{\tau}$, $\tilde{n}_{\bar{\tau}}$, and $m_{\bar{\tau}}$ via Eq. (\ref{calculate_n}) and Eq. (\ref{m_defination}). Note that parameter $\tilde{n}_{\tau}$ should be calculated via an integral over the momentum space. And other parameters are not generated from integrals but from $\tilde{n}_{\tau}$ directly (see Eq. (\ref{m_defination})). $\tilde{n}_{\tau}$ is determined by the relative position of the energy with respect to the Fermi level (or the chemical potential) (see appendix C). Hence, we neglect constant terms in the mean-field process which shift all energy bands equally and have no effect on the integral of $\tilde{n}_{\tau}$ and the self-consistent process. The updated $(\tilde{n}_{\tau}, m_{\tau}, \tilde{n}_{\bar{\tau}}, m_{\bar{\tau}})$ corresponds to a new point in the parameter space, denoted by a vector $\textbf{P}_{\text{update}}$. We define the distance of the two points as the deviation
\begin{equation}
\delta = |\textbf{P}_{\text{given}}-\textbf{P}_{\text{update}}|.
\end{equation}
For a given point in the parameter space, if it is a solution of the set of mean field equations (MFEs), then $\delta$ is zero. We thus scan the entire parameters space and try to find the parameter vectors where $\delta$ converges to zero. And we define these parameter vectors as the solutions.

\begin{figure}[t!]
	\begin{center}	
\includegraphics[width=1.0\columnwidth]{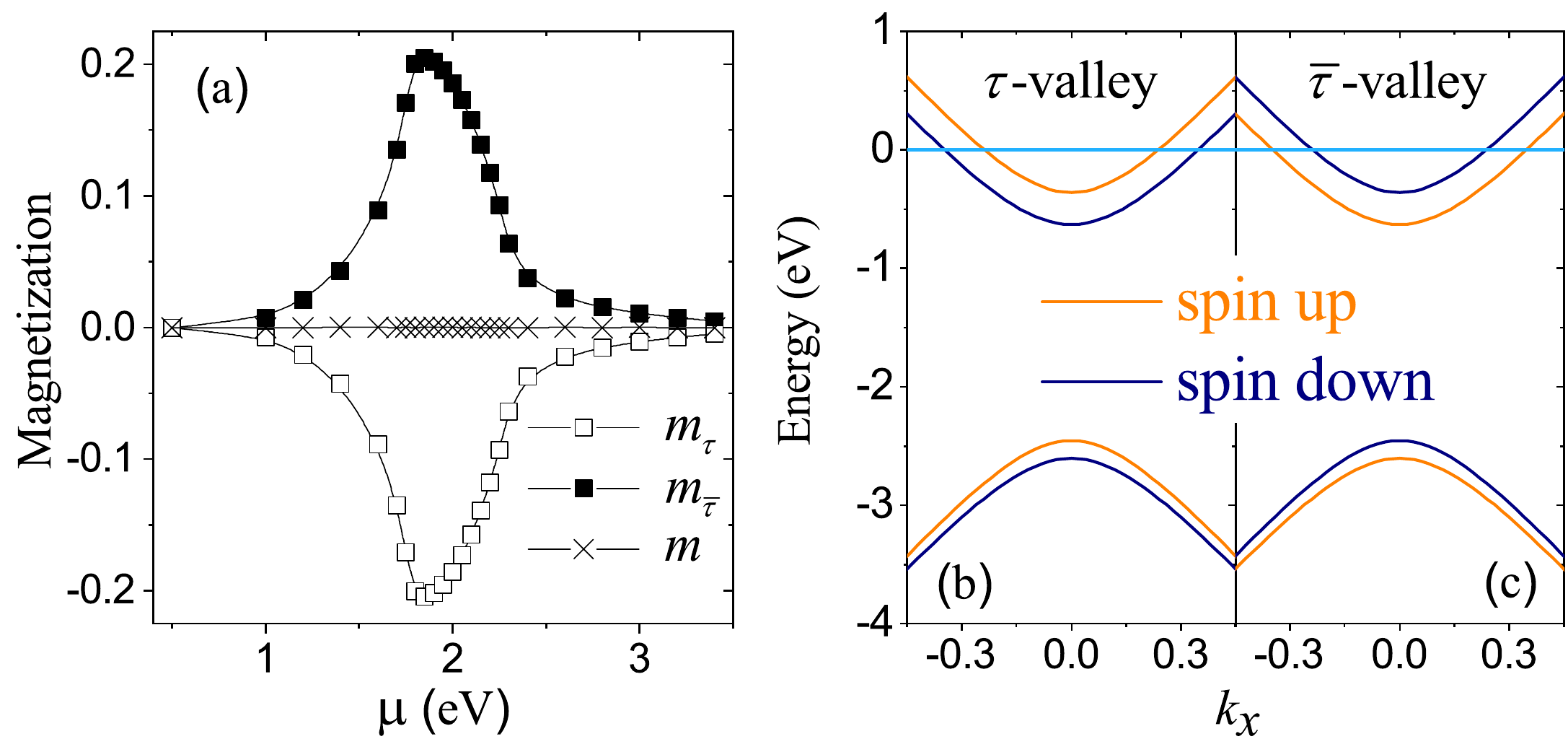}
	\end{center}
	\caption{(Color online) (a) The dependence of $m_{\tau}$, $m_{\bar{\tau}}$ and $m$ on $\mu$. $U=1.0$ eV, $U^{\prime} = 0$ eV. (b) and (c) The energy spectrum $\mathbb{E}_{\text{c}\tau s}(\bm{k})$ along $k_{x}$ direction.  $U = 1.0$ eV, $U^{\prime} = 0$ eV and $\mu =1.7$ eV. The horizontal line shows the Fermi surface.
 }   \label{Fig3}
\end{figure}

It is found that the solution of MFEs is not unique (see Fig.~\ref{solutions_of_MFEs}). And the solutions are characterized by converged parameter vectors in the parameter space. In the numerical calculation, we grid the definitional domain of each parameters into $\mathcal{N}$ subintervals, that is, the parameter space is grided into $\mathcal{N}^4$ subspaces. As $\mathcal{N}$ goes to infinite, the parameter space is ergodic exactly. In practice, we take a finite $\mathcal{N}$, and $\delta$ is kept at the order $10^{-6}$. In this paper, we take $a = 3.193$ $\text{\AA}$, $t = 1.1$ eV, $\Delta = 1.66$ eV, and $2\lambda = 0.15$ eV, which are the fitting results to the ab initio calculation~\cite{Xiao2012}. The intravalley Coulomb interaction $U$ is usually unknown in TMDCs. According to the discussion of R. Rold\'an {\it{et al}}.~\cite{Roldan2013}, electronic states in the neighbor region of $\textbf{K}$ and $\textbf{K}^{\prime}$ points are characterized by the $4d$ orbitals of Mo atoms. The order of magnitude of $U_{4d}$ is approximated by the ionization energy of Mo atom. According to previous investigations of $\text{MoS}_{2}$~\cite{Roldan2013,Rostami2015}, one usually takes $U_{4d}\approx 2.0 \sim 4.0$ eV. We compare our definition of the interacting Hamiltonian with that of Rostami~\cite{Rostami2015}, we find $U = U_{4d}/2$. Therefore, the intravalley Coulomb interaction is about $U \approx 1.0 \sim 2.0$ eV. Basing on the above consideration, we take $U=1.0$ eV and find the result is reasonable, when $U$ is combined with other parameter values.

Fig.~\ref{solutions_of_MFEs} shows the evolution of $\text{E}_{\text{free}}$ and $\delta$ with $\tilde{n}_{\tau}$, $m_{\tau}$, $\tilde{n}_{\bar{\tau}}$, and $m_{\bar{\tau}}$ in (a)-(d), respectively ($\mathcal{N}=150$). We obtain three solutions: one solution $L_{0}$ and two degenerated solutions $L_{\pm}$ (with the same free energy). $\text{E}_{\text{free}}$ is not the total free energy because the constant terms have been neglected in the calculation (see Eq.(\ref{B1}) and Eq. (\ref{B6})). As for different mean field solutions ($L_{0}$ and $L_{\pm}$), the values of the neglected constant are different. In Fig.2, we show $\text{E}_{\text{free}}$  of $L_{0}$ and $L_{\pm}$ on the same energy scale for convenience without the meaning of comparison.  $L_{\pm}$ are new solutions, which haven't been obtained previously \textit{due to the ignoring of the intervalley Coulomb interaction} \cite{Donck2018,Braz2018}. In Fig. \ref{solutions_of_MFEs}(a) and (c),  $\tilde{n}_{\tau}(L_{+})$ and $\tilde{n}_{\bar{\tau}}(L_{+})$ are very close  (the same to the solution of $L_{-}$ ).
At the numerical precision $\delta\approx10^{-6}$, we obtain the difference $\tilde{n}_{\tau}(L_{+})-\tilde{n}_{\bar{\tau}}(L_{+})$  is not zero but about the order of $10^{-4}$ indicating a slight valley polarization.
However, $\tilde{n}_{\tau}=\tilde{n}_{\bar{\tau}}$ for $L_{0}$ (see Fig.~\ref{solutions_of_MFEs}(a) and (c)).
As for $L_{0}$, the states of $\tau$ and $\bar{\tau}$-valley can be spin polarized but in opposite directions, which contributes a zero net magnetization (Fig. \ref{solutions_of_MFEs} (b) and (d)).
In contrast, for solutions $L_{+}$ and $L_{-}$, spin polarization for both valleys can be induced as well but in the same direction, leading to a net magnetization for each solution.
Because $L_{\pm}$ are two degenerated solutions, the spin-polarized states (composed of two valleys) from  $L_{+}$ and  $L_{-}$ are aligned opposite,
giving rise to a zero net magnetization since the state of entire system is randomly composed of the states of $L_{+}$ and  $L_{-}$~\cite{Roch2020}.
We further speculate that the states of $L_{\pm}$ may manifest themselves by forming spin-polarized ``domains"  in real materials. And globally there is no net magnetization without introducing an external magnetic field.

We firstly discuss $U^{\prime}=0$ case, and the effective mean field becomes
$\mathbb{F}_{\tau s} =  \frac{2}{N}
\sum_{\bm{k}}U\langle n^{\tau }_{\bm{k} \bar{s}}\rangle$.
We derive the solution $L_{0}$ that satisfies
$\sum_{\bm{k}}\langle n^{\tau }_{\bm{k} \uparrow}\rangle=
\sum_{\bm{k}}\langle n^{\bar{\tau} }_{\bm{k} \downarrow}\rangle$, and
$\sum_{\bm{k}}\langle n^{\tau }_{\bm{k} \downarrow}\rangle=
\sum_{\bm{k}}\langle n^{\bar{\tau} }_{\bm{k} \uparrow}\rangle$.
These indicate that the spin splitting of the conduction band at $\tau$ and $\bar{\tau}$ is inverted due to time-reversal symmetry (TRS) (see Fig.~\ref{Fig3}(b) and (c))~\cite{Xiao2012}. It can be seen from energy gap in Eq. (\ref{egap}) at the minimum of the conduction band ($k = 0$), $\Delta^{\text{c}\tau}_{E} \left( 0\right) = - 2 U m_{\tau}$. From Fig.~\ref{Fig3}(a), we find that $m_{\tau} = - m_{\bar{\tau}}$, and then $\Delta^{\text{c}\tau}_{E}(0)$ always takes the opposite values
(see Fig.~\ref{Fig3}(b) and (c)). We only derive the $L_{0}$ solution in this case, which means an unpolarized state in absence of intervalley interaction.

\begin{figure}[tb]
	\begin{center} \includegraphics[width=1.0\columnwidth]{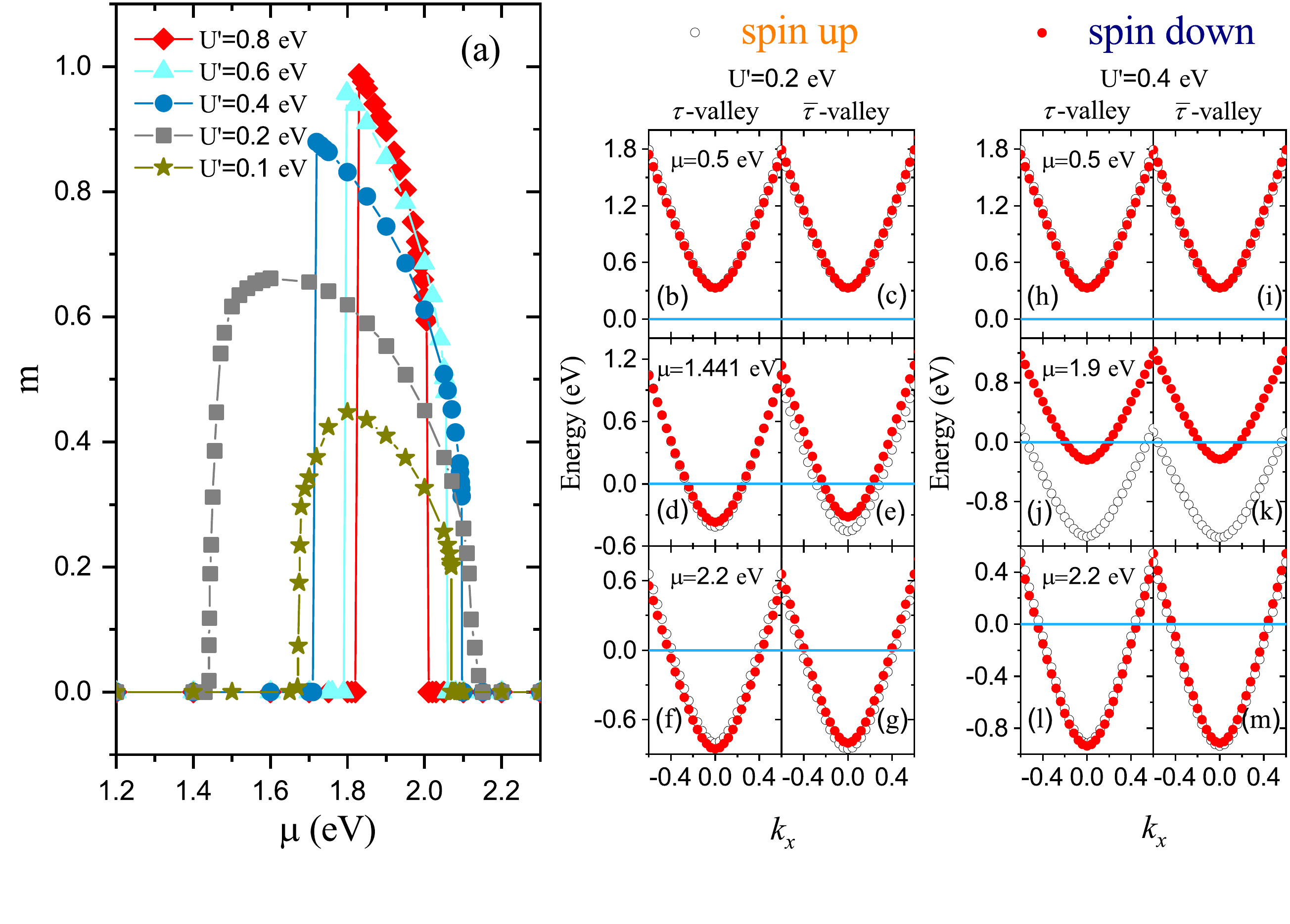}
	\end{center}
	\caption{(Color online) (a) The dependence of $m$ on $\mu$ at various $U^{\prime}$. With the increase of $\mu$, the solution of MFEs changes between $L_{0}$ and $L_{\pm}$. When the solutions are $L_{\pm}$, we merely chose $L_{+}$ and calculate $m$ of $L_{+}$, because $L_{+}$ and $L_{-}$ are degenerate.
(b)-(m) The spin splitting of the conduction band along the $k_{x}$ direction. The horizontal lines represent the Fermi level. $U = 1.0$ eV.
} \label{Fig4}
\end{figure}

In general, the intervalley e-e interaction is comparable to the intravalley interaction even at high electron density, due to a small Bohr radius $a_{B}\sim 0.5 $ nm~\cite{Dery2016}. In this case, all of  $\tilde{n}_{\tau}$, $m_{\tau}$, $\tilde{n}_{\bar{\tau}}$, and $m_{\bar{\tau}}$ appear in the mean field $\mathbb{F}_{\tau \text{s}}$ (Eq. (\ref{MF_total})), i.e. carriers in the two valleys are interacted with each other. Fig.~\ref{Fig4}(a) shows the dependence of $m$ on $\mu$ at various $U^{\prime}$ for $L_{+}$ state. The main feature of the $L_{+}$  state is that there is a region of gate voltage (characterized by chemical potential $\mu$) in which a net magnetization is developed and characterized by a finite $m$. However, the global net magnetization is zero due to the superposition of the $L_{\pm}$ states. We may call it as a ferromagnetic (FM) state. Out of this region, $m=0$ means a paramagnetic (PM) state. Therefore, there are two borders between the FM and PM state located at a lower and a higher $\mu$, which indicate a PM-FM phase transition and vice versa. However, three kinds of transitions are found. The first one is shown for $U^{\prime}=0.2$ eV, where $m$ grows and disappears with $\mu$ continuously, indicating a second-order phase transition at two borders; the second one is shown for $U^{\prime}=0.4$, $0.6$, and $0.8$ eV, where the PM-FM transitions are emergent discontinuously, which indicate a first-order phase transition and consistent with the experimental observations \cite{Roch2020}. The third one ($U^{\prime}=0.1$ eV) shows a second-order and a first-order at the lower and higher $\mu$, respectively. As a theoretical investigation, we study the effects due to various parameters to cover most possibilities. The validity of the parameters should get supports from experimental observations or other ways. Our results show that when $U^{\prime}\ge 0.4$ eV, the phase transition is clearly of first order, which agrees with the experimental results ~\cite{Roch2020}. This is also consistent with the previous prediction that the intervalley interaction is comparable with the intravalley interaction even at high electron density~\cite{Dery2016}. We therefore deduce that realistic intervalley Coulomb interaction should be in this range. The complicated transition behaviors exhibited in other parameter ranges might not be a reality. It is still lack of an intuitive picture for appearance of such a complicated case.

To understand the existence of the FM state, we show band structures in Figs.~\ref{Fig4}(b)-(m) and the relative positions of Fermi level to CBM. The electron Coulomb interaction renormalizes the band structures (or the position of CBM). For different $U^{\prime}$, the relative position of the Fermi level (or chemical potential) to the CBM is different. The PM states at small $\mu$ can be understood because the Fermi level does not pass through any bands (Figs. \ref{Fig4}(b), (c), (h) and (i)). For the PM states at large $\mu$ (Figs. \ref{Fig4}(f), (g), (l) and (m)), the Fermi level deeply lies in all four bands where the e-e interaction may be weak due to a high electron density, and the spin splitting due to the e-e interaction is insignificant. In contrast to these two cases, Figs. \ref{Fig4}(d), (e), (j) and (k) show the band structures for the FM states. It is noted that the spin splitting of bands is obviously observed and the Fermi level is not deeply lying in the conduction bands, but lies just around the bottom of some bands. This result matches our intuitive picture that the electron Coulomb interaction should be more important when the Fermi level is close to the CBM.~\cite{Roch2019} For Figs. \ref{Fig4}(d) and (e), it seems a ``normal" FM state in which the Fermi level is not far away from the bottom of four bands. However, for Figs. \ref{Fig4}(j) and (k), the Fermi level is deeply in the spin-up bands but shallowly lies in the spin-down bands. It might be this difference leading to a different transition order between the PM and FM existing at lower and higher $\mu$, respectively. It is obvious that the relative position of the Fermi level and the CBM is quite crucial for existence of the FM states. And this relative position is altered dramatically by including the electron Coulomb interaction and can not qualitatively predicted by thinking about the picture of non-interaction case.

The complicate behaviors of the transition induced by $\mu$ may rest themselves into a fact that the energy bands are altered in the self-consistent MFEs. Comparable to the experiments, it seems that the first-order phase transitions at the two borders may be consistent with experimental observations \cite{Roch2020}. If this is the case, we can deduce that the $U^{\prime}$ may be in the range of $0.4$-$0.8$ eV. This energy scale may be converted to a length scale which corresponds to a Coulomb length for $U^{\prime}$ and a size of the so-called ``puddle" in experiments, which is in $1$-$2$ nm. This can be tested in experiments although this size is said to be small but not given in experiments. So far, we know that the polarized ``puddles" resemble the domains in usual ferromagnets. In zero-$\mathbf{B}$ case, the polarizations of these ``puddles" may be randomly distributed giving rise to a zero net magnetization. We may speculate that the polarizations of ``puddles" may be aligned into one direction when applying a nonzero $\mathbf{B}$, which bring us a net magnetization. This scenario is consistent with the experimental observation \cite{Roch2019}.  A further measurement on this size can demonstrate our theory clearly. We should emphasize that an FM state can be induced by tuning gate voltage due to finite $U^{\prime}$. This reflects the important role of intervalley Coulomb interaction. The FM state can be derived only in the presence of  intervalley Coulomb interaction. In the absence of the Coulomb interaction, Eq.(2) shows that the ground state is both valley and spin degenerate at k=0. The conduction band at $\tau$ and $\bar{\tau}$ is inverted as the requirement of TRS. However, in the presence of the Coulomb interaction, it is found that the valley and spin degeneracy are lifted (Fig. 4(d), (e), (j), (k)). Note that the energy difference with the same spin index and various valley index in Fig. 4(j) and (k) is so small $|\mathbb{E}_{\text{c}\tau s} \left( 0 \right) - \mathbb{E}_{\text{c} \bar{\tau} s} \left( 0 \right)| \approx 0.01$ eV that it is hard to be recognized. As shown in Fig. 4(d), (e), (j) and (k), intervalley Coulomb interaction combined with suitable Fermi level induce the spin polarization of the ground state, which doesn’t satisfy the requirement of TRS. Therefore, the TRS can be broken by the joint effects of intervalley Coulomb interaction and the Fermi level. A slight valley polarization (imbalanced of electrons distribution at $\tau$ and $\bar{\tau}$ valley)~\cite{Zeng2012,Song2017} can be induced by the e-e interaction at $10^{-3}$ order (see Fig.~\ref{Fig4}(j),~(k)). When $\mu$ increases further, electron density is increased, e-e interaction is reduced, valley degeneracy and the TRS recovers again (see Fig.~\ref{Fig4}(l) and (m)).

This work is supported in part by the National Key R\&D Program of China (Grant No. 2018YFA0305800), the NSFC (Grant Nos. 11974348, 11674317, and 11834014). It is also supported by the Fundamental Research Funds for the Central Universities, and the Strategic Priority Research Program of CAS (Grant Nos. XDB28000000, and XDB33000000).

\appendix

\section{Model}

\subsection{Solve noninteracting Hamiltonian}

According to the work reported by Xiao Di \emph{et al.}~\cite{Xiao2012}, the effective Hamiltonian of ML-$\text{MoS}_{2}$ around Dirac cones without Coulomb interaction is
\begin{equation} \label{1}
\hat{\text{H}}_{0} = at\left(\tau k_{x} \sigma_{x} + k_{y} \sigma_{y}\right) + \frac{ \Delta}{2}\sigma_{z} - \lambda\tau\frac{\sigma_{z} - 1}{2} \hat{s}_{z},
\end{equation}
where $\tau = \pm 1$ is the valley index. The spin splitting caused by spin orbital coupling is $2\lambda$. $\sigma_{\alpha}$ ($\alpha = x, y, z$) is the Pauli matrices. $a$ is the lattice constant. $t$ is the hopping integral. $\Delta$ is the energy gap between the conduction band and the valence band (when $\lambda = 0$). $\hat{s}_{z}$ is the $z$ component of the spin operator. For convenience, we choose a diamond Brillouin zone (BZ) in the following calculation (see Fig.~\ref{FigS2}). We explicitly write
\begin{equation} \label{2}
\hat{\bm{s}}_{\bm{z}} = \left(
\begin{array} {cc}
s_{z_{1}} & 0    \\
0      & s_{z_{2}}
\end{array}
\right),\,\,\,\,\,\,\,\,\,\,
\bm{\tau} = \left(
\begin{array} {cc}
\tau_{1} & 0    \\
0        & \tau_{2}
\end{array}
\right),\,\,\,\,\,\,\,\,\,\,
\end{equation}
where $s_{z_{1}}=1$ and $s_{z_{2}}=-1$ represent spin up and spin down respectively, $\tau_{1}=1$ and $\tau_{2}=-1$ indicate the two valleys located at $\textbf{K}$ and $\textbf{K}^{\prime}$. We perform direct product for the valley, spin and band (conduction band and valence band) index freedom in the Hamiltonian. $\hat{\text{H}}_{0}$ is rewritten as
\begin{eqnarray} \label{3}
\hat{\text{H}}_{0} &=& at \left(\bm{\tau}\otimes k_{x} \sigma_{x} \otimes \bm{1} + \bm{1}\otimes k_{y} \sigma_{y} \otimes \bm{1}\right) + \bm{1}\otimes\frac{ \Delta}{2} \sigma_{z} \otimes \bm{1}   \notag\\
&-& \lambda\bm{\tau}\otimes\frac{\sigma_{z} - 1}{2} \otimes\hat{\bm{s}}_{\bm{z}},
\end{eqnarray}
where $\bm{1}$ is the identity matrix, and $\otimes$ denotes direct product. It is obvious that $\hat{\text{H}}_{0}$ is a $8\times8$ matrix.
Substituting the Pauli matrix,
\begin{equation} \label{4}
\sigma_{x}=\left(
\begin{array}{cc}
0  &  1\\
1  &  0
\end{array}
\right),\
\sigma_{y}=\left(
\begin{array}{cc}
0  &  -i\\
i  &  0
\end{array}
\right),\
\sigma_{z}=\left(
\begin{array}{cc}
1  &  0\\
0  &  -1
\end{array}
\right),\
\end{equation}
into Eq.(\ref{3}), one obtains
\begin{equation} \label{5}
\hat{\text{H}}_{0} = \left(
\begin{array}{cc}
\bm{\alpha} & \bm{0} \\
\bm{0}      & \bm{\beta}
\end{array}
\right).
\end{equation}
$\hat{\text{H}}_{0}$ is a block matrix, where $\bm{0}$, $\bm{\alpha}$ and $\bm{\beta}$  are $4\times4$ matrixes. $\bm{0}$ is a zero matrix,
\begin{equation} \label{6}
\bm{\alpha} = \left(
\begin{array}{cccc}
\frac{\Delta}{2} & 0 & \alpha_{-} & 0 \\
0 & \frac{\Delta}{2} & 0 & \alpha_{-} \\
\alpha_{+} & 0 & -\frac{\Delta}{2} + \lambda\tau_{1} s_{z_{1}} & 0 \\
0 & \alpha_{+} & 0 & -\frac{\Delta}{2} + \lambda\tau_{1} s_{z_{2}}
\end{array}
\right),
\end{equation}
where $\alpha_{\pm}=at(\tau_{1} k_{x}\pm ik_{y})$, and
\begin{equation}
\bm{\beta} = \left(
\begin{array}{cccc}
\frac{\Delta}{2} & 0 & \beta_{-} & 0 \\
0 & \frac{\Delta}{2} & 0 & \beta_{-} \\
\beta_{+} & 0 & -\frac{\Delta}{2} + \lambda\tau_{2} s_{z_{1}} & 0 \\
0 & \beta_{+} & 0 & -\frac{\Delta}{2} + \lambda \tau_{2} s_{z_{2}}
\end{array}
\right),
\end{equation}
where $\beta_{\pm}=at(\tau_{2} k_{x} \pm i k_{y})$.
It is easy to diagonalize $\bm{\alpha}$ matrix. The energy eigenvalues for $\tau_{1}$ valley reads
\begin{align}
E^{(1)}_{\tau_{1} s_{z_{1}}} &= \frac{\lambda\tau_{1} s_{z_{1}}}{2} - \sqrt{\left(atk\right)^{2} + \left(\frac{\Delta - \lambda\tau_{1} s_{z_{1}}}{2} \right)^{2}}, \\
E^{(2)}_{\tau_{1}s_{z_{1}}} &= \frac{\lambda\tau_{1}s_{z_{1}}}{2} + \sqrt{\left(atk\right)^{2} + \left(\frac{\Delta - \lambda\tau_{1} s_{z_{1}}}{2} \right)^{2}}, \\
E^{(3)}_{\tau_{1}s_{z_{2}}} &= \frac{\lambda\tau_{1}s_{z_{2}}}{2} - \sqrt{\left(atk\right)^{2} + \left(\frac{\Delta - \lambda\tau_{1} s_{z_{2}}}{2} \right)^{2}}, \\
E^{(4)}_{\tau_{1}s_{z_{2}}} &= \frac{\lambda\tau_{1}s_{z_{2}}}{2} + \sqrt{\left(atk\right)^{2} + \left(\frac{\Delta - \lambda\tau_{1}s_{z_{2}}}{2} \right)^{2}}.
\end{align}
The corresponding eigenvectors are
\begin{eqnarray}
\bm{u}^{(1)}_{\tau_{1}s_{z_{1}}} &=& \text{N}^{(1)}_{\tau_{1}s_{z_{1}}} \left(  \frac{\Delta - 2E^{(2)}_{\tau_{1}s_{z_{1}}}}{2at(\tau_{1}k_{x} + ik_{y})}, 0,1,0 \right)^{\text{T}}, \\
\bm{u}^{(2)}_{\tau_{1}s_{z_{1}}} &=&  \text{N}^{(2)}_{\tau_{1}s_{z_{1}}}\left(\frac{\Delta - 2E^{(1)}_{\tau_{1}s_{z_{1}}}}{2at(\tau_{1}k_{x} + ik_{y})}, 0,1,0 \right)^{\text{T}}, \\
\bm{u}^{(3)}_{\tau_{1}s_{z_{2}}} &=& \text{N}^{(3)}_{\tau_{1}s_{z_{2}}} \left(0,  \frac{\Delta - 2E^{(4)}_{\tau_{1}s_{z_{2}}}}{2at(\tau_{1}k_{x} + i k_{y})}, 0,1 \right)^{\text{T}}, \\
\bm{u}^{(4)}_{\tau_{1}s_{z_{2}}} &=& \text{N}^{(4)}_{\tau_{1}\text{s}_{z_{2}}} \left(0,  \frac{\Delta - 2E^{(3)}_{\tau_{1}s_{z_{2}}}}{2at(\tau_{1}k_{x} + ik_{y})}, 0,1 \right)^{\text{T}},
\end{eqnarray}
\begin{figure}[t!]
	\begin{center}
		\includegraphics[width=\columnwidth]{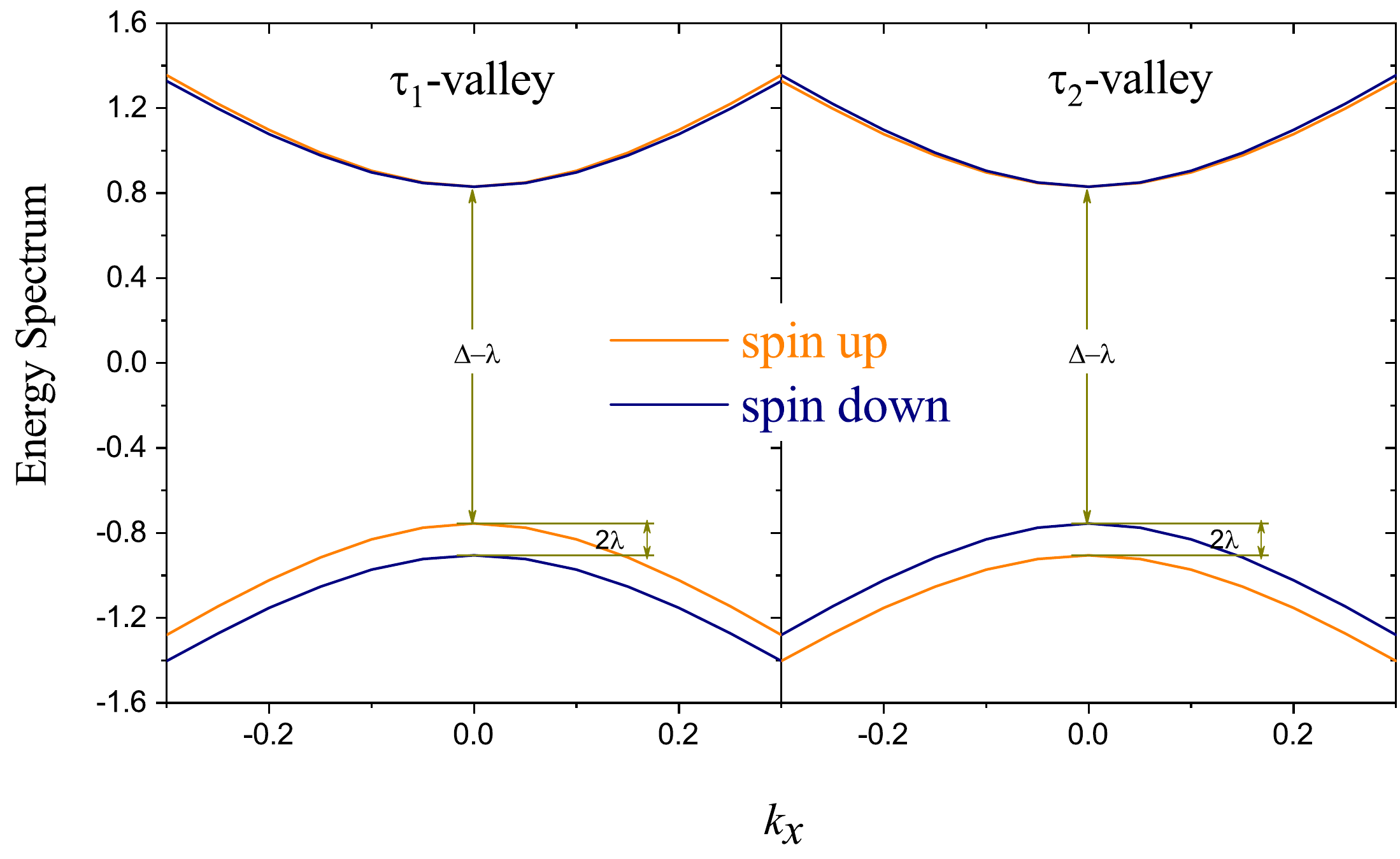}
	\end{center}
	\caption{(Color online) Energy spectrum along $k_{x}$ direction at $\tau_{1}$ (Left) and $\tau_{2}$ (Right) valley. $\text{a} = 3.193~\text{\AA}$, $\text{t} = 1.1$ eV, $\Delta = 1.66$ eV, and $2\lambda = 0.15$ eV. }   \label{FigS1}
\end{figure}
where the eigenvectors are normalized by
\begin{align}
\text{N}^{(1)}_{\tau_{1}s_{z_{1}}} &= \sqrt{\frac{4(atk)^{2}}{(\Delta - 2E^{(2)}_{\tau_{1}s_{z_{1}}})^{2} + 4(atk)^{2}}}, \\
\text{N}^{(2)}_{\tau_{1}s_{z_{1}}}  &= \sqrt{\frac{4(atk)^{2}}{(\Delta - 2E^{(1)}_{\tau_{1}s_{z_{1}}})^{2} + 4(atk)^{2}}}, \\
\text{N}^{(3)}_{\tau_{1}s_{z_{2}}} &= \sqrt{\frac{4(atk)^{2}}{(\Delta - 2E^{(4)}_{\tau_{1}s_{z_{2}}})^{2} + 4(atk)^{2}}}, \\
\text{N}^{(4)}_{\tau_{1}s_{z_{2}}}  &= \sqrt{\frac{4(atk)^{2}}{(\Delta - 2E^{(3)}_{\tau_{1}s_{z_{2}}})^{2} + 4(atk)^{2}}}.
\end{align}
In the same way, we diagonalize $\bm{\beta}$ matrix obtaining the energy eigenvalues at $\tau_{2}$ valley
\begin{align}
E^{(5)}_{\tau_{2}s_{z_{1}}} &= \frac{\lambda\tau_{2}s_{z_{1}}}{2} - \sqrt{\left(atk\right)^{2} + \left(\frac{\Delta - \lambda\tau_{2}s_{z_{1}}}{2} \right)^{2}}, \\
E^{(6)}_{\tau_{2}s_{z_{1}}} &= \frac{\lambda\tau_{2}s_{z_{1}}}{2} + \sqrt{\left(atk\right)^{2} + \left(\frac{\Delta - \lambda\tau_{2}s_{z_{1}}}{2} \right)^{2}},
\end{align}
\begin{align}
E^{(7)}_{\tau_{2}s_{z_{2}}} &= \frac{\lambda\tau_{2}s_{z_{2}}}{2} - \sqrt{\left(atk\right)^{2} + \left(\frac{\Delta - \lambda\tau_{2}s_{z_{2}}}{2} \right)^{2}}, \\
E^{(8)}_{\tau_{2}s_{z_{2}}} &= \frac{\lambda\tau_{2}s_{z_{2}}}{2} + \sqrt{\left(atk\right)^{2} + \left(\frac{\Delta - \lambda\tau_{2}s_{z_{2}}}{2} \right)^{2}}.
\end{align}
The eigenvectors are
\begin{align}
\bm{u}^{(5)}_{\tau_{2}s_{z_{1}}} &= \text{N}^{(5)}_{\tau_{2}s_{z_{1}}} \left(  \frac{\Delta - 2E^{(6)}_{\tau_{2}s_{z_{1}}}}{2at(\tau_{2}k_{x} + ik_{y})}, 0,1,0 \right)^{\text{T}}, \\
\bm{u}^{(6)}_{\tau_{2}s_{z_{1}}} &= \text{N}^{(6)}_{\tau_{2}s_{z_{1}}} \left(  \frac{\Delta - 2E^{(5)}_{\tau_{2}s_{z_{1}}}}{2at(\tau_{2}k_{x} + ik_{y})}, 0,1,0 \right)^{\text{T}}, \\
\bm{u}^{(7)}_{\tau_{2}s_{z_{2}}} &= \text{N}^{(7)}_{\tau_{2}s_{z_{2}}} \left(0,  \frac{\Delta - 2E^{(8)}_{\tau_{2}s_{z_{2}}}}{2at(\tau_{2}k_{x} + ik_{y})},0,1 \right)^{\text{T}}, \\
\bm{u}^{(8)}_{\tau_{2}s_{z_{2}}} &= \text{N}^{(8)}_{\tau_{2}s_{z_{2}}} \left(0,  \frac{\Delta - 2E^{(7)}_{\tau_{2}s_{z_{2}}}}{2at(\tau_{2}k_{x} + ik_{y})}, 0,1 \right)^{\text{T}},
\end{align}
where
\begin{align}
\text{N}^{(5)}_{\tau_{2}s_{z_{1}}} &= \sqrt{\frac{4(atk)^{2}}{(\Delta - 2E^{(6)}_{\tau_{2}s_{z_{1}}})^{2} + 4(atk)^{2}}},\\
\text{N}^{(6)}_{\tau_{2}s_{z_{1}}} &= \sqrt{\frac{4(atk)^{2}}{(\Delta - 2E^{(5)}_{\tau_{2}s_{z_{1}}})^{2} + 4(atk)^{2}}},\\
\text{N}^{(7)}_{\tau_{2}s_{z_{2}}} &= \sqrt{\frac{4(atk)^{2}}{(\Delta - 2E^{(8)}_{\tau_{2}s_{z_{2}}})^{2} + 4(atk)^{2}}},\\
\text{N}^{(8)}_{\tau_{2}s_{z_{2}}} &= \sqrt{\frac{4(atk)^{2}}{(\Delta - 2E^{(7)}_{\tau_{2}s_{z_{2}}})^{2} + 4(atk)^{2}}}.
\end{align}
Energy eigenvalues are written compactly as~\cite{Yu2015}
\begin{equation} \label{energy_spectrum_H0}
E_{n \tau s} = \frac{1}{2}\lambda\tau s \pm \sqrt{(atk)^{2} + (\frac{\Delta - \lambda\tau s}{2} )^{2}},
\end{equation}
where the spin index $s=s_{z1}$ or $s_{z2}$. $E_{n \tau s}$ is shown in Fig.~\ref{FigS1}. The up plus sign denotes the conduction band (c). The bottom minus sign denotes the valence band (v). $n$ is the band index (conduction band $n = c$, valence band $n = v$). $k$ is the module of the wave vector, $k=\sqrt{k^{2}_{x} + k^{2}_{y}}$. The valley index $\tau=\tau_{1}$ or $\tau_{2}$.  The corresponding eigenstate is a superposition state of the bases~\cite{Xiao2012} with the coefficients defined by the eigenvectors, which is denoted as $|n\tau k s\rangle=\psi_{n\tau s}(\bm{k}, \bm{r})$.

\subsection{Coulomb interaction}

Electron-electron (e-e) interactions have significant effects on the physical properties of monolayer materials~\cite{Chernikov2014}. As early as in 1979, Keldysh investigated Coulomb interaction in thin semiconductor and semimetal films, and gave an effective Coulomb interaction, which  is expressed by the Neumann and Struve functions  ~\cite{Keldysh1979}. In this paper, we focused on a qualitative discussion. Therefore, we take the usual bare Coulomb interaction, instead of the complicated potential given by Keldysh. The bare Coulomb interaction is
\begin{equation}
\text{V}(\bm{r}_{1} - \bm{r}_{2}) = \frac{e^2}{4\pi\epsilon_{0}} \frac{1}{|\bm{r}_{1} - \bm{r}_{2}|},
\end{equation}
where $e$ is the elementary charge, $\epsilon_{0}$ is the vacuum permittivity. It is obvious that in terms of field operators the Coulomb interaction is written as~\cite{Bruus2004}
\begin{align}
\hat{\text{V}}(\bm{r}_{1} - \bm{r}_{2}) =& \frac{1}{2}\sum_{s_{1}s_{2}} \iint d\bm{r}_{1}d\bm{r}_{2}\text{V}(\bm{r}_{1} - \bm{r}_{2}) \nonumber\\ &\psi^{\dagger}_{s_{1}}(\bm{r}_{1})\psi^{\dagger}_{s_{2}}(\bm{r}_{2})
\psi_{s_{2}}(\bm{r}_{2})\psi_{s_{1}}(\bm{r}_{1}).
\end{align}
We take the transformation
\begin{align}
\psi^{\dagger}_{s}(\bm{r}) &= \frac{1}{\sqrt{N}} \sum_{n\tau\bm{k}} \psi^{\ast}_{n\tau s}(\bm{k}, \bm{r}) a^{n\tau\dagger}_{\bm{k}s},\\
\psi_{s}(\bm{r}) &= \frac{1}{\sqrt{N}} \sum_{n\tau\bm{k}} \psi_{n \tau \text{s}}(\bm{k}, \bm{r}) a^{n\tau}_{ \bm{k} s}.
\end{align}
It is secondly quantized in the $| n \tau \bm{k} s \rangle$ representation~\cite{Bruus2004}
\begin{equation} \label{v_definition}
\hat{\text{V}} = \frac{1}{N} \sum_{s_{1}s_{2}} \sum_{\substack{\bm{k}_{1}\bm{k}_{2}\\ \bm{k}_{3} \bm{k}_{4}}} \sum_{\substack{n_{1}n_{2} \\ n_{3} n_{4}}}
\sum_{\substack{\tau_{1}\tau_{2} \\ \tau_{3} \tau_{4}}} \text{V}_{\text{int}} a^{n_{1}\tau_{1}\dagger}_{\bm{k}_{1} s_{1}} a^{n_{2}\tau_{2}\dagger}_{\bm{k}_{2} s_{2}}
a^{n_{3}\tau_{3}}_{\bm{k}_{3} s_{2}}
a^{n_{4}\tau_{4}}_{\bm{k}_{4} s_{1}},
\end{equation}
where $\text{V}_{\text{int}}$ denotes the strength of the e-e interaction and $N$ is the number of the unit cell. $a^{n\tau\dagger}_{\bm{k} s}$ $(a^{n \tau}_{\bm{k} s})$ is the creation (annihilation) operator at $| n \tau \bm{k} s \rangle$ state. Because the valence band is fully filled, we only consider the e-e interaction in the conduction band, i.e. $n = c$. In the following derivation, the superscript $c$ is omitted. We take the summation of $\tau_{2}$, $\tau_{3}$ and $\tau_{4}$ in Eq.(\ref{v_definition}), obtaining
\begin{equation}
\hat{\text{V}} = \frac{1}{N}
\sum_{\tau s_{1} s_{2}}
\sum_{\bm{k}_{1}\bm{k}_{2} \bm{k}_{3} \bm{k}_{4}}
\sum_{i = 1}^{8}
\text{V}^{(i)}_{\text{int}} \text{T}_{i},
\end{equation}
where $\text{T}_{i}$ reads
\begin{align}
\text{T}_{1} &=
a^{\tau\dagger}_{\bm{k}_{1} s_{1}}
a^{\tau\dagger}_{\bm{k}_{2} s_{2}}
a^{\tau}_{\bm{k}_{3} s_{2}}
a^{\tau}_{\bm{k}_{4} s_{1}}, \\
\text{T}_{2} &=
a^{\tau \dagger}_{\bm{k}_{1} s_{1}}
a^{\tau \dagger}_{\bm{k}_{2} s_{2}}
a^{\tau }_{\bm{k}_{3} s_{2}}
a^{\bar{\tau}}_{\bar{\bm{k}}_{4} s_{1}}, \\
\text{T}_{3} &=
a^{\tau\dagger}_{\bm{k}_{1} s_{1}}
a^{\tau\dagger}_{\bm{k}_{2} s_{2}}
a^{\bar{\tau} }_{\bar{\bm{k}}_{3} s_{2}}
a^{\tau}_{\bm{k}_{4} s_{1}}, \\
\text{T}_{4} &=
a^{\tau \dagger}_{\bm{k}_{1} s_{1}}
a^{\tau \dagger}_{\bm{k}_{2} s_{2}}
a^{\bar{\tau}}_{\bar{\bm{k}}_{3} s_{2}}
a^{\bar{\tau}}_{\bar{\bm{k}}_{4} s_{1}}, \\
\text{T}_{5} &=
a^{\tau \dagger}_{\bm{k}_{1}s_{1}}
a^{\bar{\tau} \dagger}_{\bar{\bm{k}}_{2}s_{2}}
a^{\tau}_{\bm{k}_{3}s_{2}}
a^{\tau}_{\bm{k}_{4}s_{1}}, \\
\text{T}_{6} &=
a^{\tau \dagger}_{\bm{k}_{1} s_{1}}
a^{\bar{\tau} \dagger}_{\bar{\bm{k}}_{2} s_{2}}
a^{\tau}_{\bm{k}_{3} s_{2}}
a^{\bar{\tau}}_{\bar{\bm{k}}_{4} s_{1}}, \\
\text{T}_{7} &=
a^{\tau \dagger}_{\bm{k}_{1}s_{1}}
a^{\bar{\tau} \dagger}_{\bar{\bm{k}}_{2}s_{2}}
a^{\bar{\tau} }_{\bar{\bm{k}}_{3}s_{2}}
a^{\tau}_{\bm{k}_{4} s_{1}}, \\
\text{T}_{8} &=
a^{\tau \dagger}_{\bm{k}_{1} s_{1}}
a^{\bar{\tau} \dagger}_{\bar{\bm{k}}_{2} s_{2}}
a^{\bar{\tau} }_{\bar{\bm{k}}_{3} s_{2}}
a^{\bar{\tau} }_{\bar{\bm{k}}_{4} s_{1}}.
\end{align}
It is obvious that $\text{T}_{1}$ gives the intravalley e-e interaction. $\text{T}_{2}$, $\text{T}_{3}$, $\text{T}_{4}$, $\text{T}_{5}$ and $\text{T}_{8}$, describe the electron transformation from one valley to the other, which are not considered in this paper. $\text{T}_{7}$ shows the intravalley transformation of electrons without exchanges of spin, which is also neglected. $\text{T}_{6}$ gives the intervalley spin exchange coupling. The interaction induced magnetic order transition of ground state is attributed to this term, which is taken into consideration carefully. The strength of the interaction corresponding to $\text{T}_{i}$ reads
\begin{align}
\text{V}^{(1)}_{\text{int}} &=
\frac{1}{2N}
\langle
\tau \bm{k}_{1},
\tau \bm{k}_{2}
| \text{V}\left( \bm{r}_{1} - \bm{r}_{2} \right) |
\tau \bm{k}_{4},
\tau \bm{k}_{3}
\rangle,\\
\text{V}^{(2)}_{\text{int}} &=
\frac{1}{2N}
\langle
\tau \bm{k}_{1},
\tau \bm{k}_{2}
| \text{V}\left( \bm{r}_{1} - \bm{r}_{2} \right) |
\bar{\tau} \bar{\bm{k}}_{4},
\tau \bm{k}_{3}
\rangle,\\
\text{V}^{(3)}_{\text{int}} &=
\frac{1}{2N}
\langle
\tau \bm{k}_{1},
\tau \bm{k}_{2}
| \text{V}\left( \bm{r}_{1} - \bm{r}_{2} \right) |
\tau \bm{k}_{4},
\bar{\tau} \bar{\bm{k}}_{3}
\rangle,\\
\text{V}^{(4)}_{\text{int}} &=
\frac{1}{2N}
\langle
\tau \bm{k}_{1},
\tau \bm{k}_{2}
| \text{V}\left( \bm{r}_{1} - \bm{r}_{2} \right) |
\bar{\tau} \bar{\bm{k}}_{4},
\bar{\tau} \bar{\bm{k}}_{3}
\rangle,\\
\text{V}^{(5)}_{\text{int}} &=
\frac{1}{2N}
\langle
\tau \bm{k}_{1},
\bar{\tau} \bar{\bm{k}}_{2}
| \text{V}\left( \bm{r}_{1} - \bm{r}_{2} \right) |
\tau \bm{k}_{4},
\tau \bm{k}_{3}
\rangle,\\
\text{V}^{(6)}_{\text{int}} &=
\frac{1}{2N}
\langle
\tau \bm{k}_{1},
\bar{\tau} \bar{\bm{k}}_{2}
| \text{V}\left( \bm{r}_{1} - \bm{r}_{2} \right) |
\bar{\tau} \bar{\bm{k}}_{4},
\tau \bm{k}_{3}
\rangle,\\
\text{V}^{(7)}_{\text{int}} &=
\frac{1}{2N}
\langle
\tau \bm{k}_{1},
\bar{\tau} \bar{\bm{k}}_{2}
| \text{V}\left( \bm{r}_{1} - \bm{r}_{2} \right) |
\tau \bm{k}_{4},
\bar{\tau} \bar{\bm{k}}_{3}
\rangle,\\
\text{V}^{(8)}_{\text{int}} &=
\frac{1}{2N}
\langle
\tau \bm{k}_{1},
\bar{\tau} \bar{\bm{k}}_{2}
| \text{V}\left( \bm{r}_{1} - \bm{r}_{2} \right) |
\bar{\tau} \bar{\bm{k}}_{4},
\bar{\tau} \bar{\bm{k}}_{3}
\rangle.
\end{align}
$\bar{\tau}$ $(\bar{s})$ represents the opposite valley (spin) of $\tau$ $(s)$. $\bm{k}$ ($\bar{\bm{k}}$) indicates the relative wave vector with respect to the minimum of $\tau$ ($\bar{\tau}$) valley.
As for $\text{T}_{1}$, we take $s_{2} = \bar{s}_{1}$. Due to the Pauli exclusion principle, electrons with the opposite spin is apt to be spatially closer than those with the same spin. Therefore, the contribution of the term $s_{2} = s_{1}$ is omitted. The momentum conservation is employed. In $\text{T}_{1}$ term, we take $\bm{k}_{4}=\bm{k}_{1}$ and $\bm{k}_{3}= \bm{k}_{2}$. As for $\text{T}_{6}$, we are focused on the spin exchange and neglect the momentum scattering in the process. Therefore, we take $\bm{k}_{3}=\bm{k}_{1}$ and $\bm{k}_{4}=\bm{k}_{2}$. It is convenient to define $U = \text{V}^{(1)}_{\text{int}}$ and $U^{\prime} = \text{V}^{(6)}_{\text{int}}$. The intravalley and intervalley e-e interaction are then written as
\begin{align}
\hat{\text{V}}_{\text{intra}} &=
\frac{1}{N}
\sum_{\tau \text{s}}
\sum_{\bm{k}_{1}\bm{k}_{2}}
U
a^{\tau\dagger}_{\bm{k}_{1} s}
a^{\tau\dagger}_{\bm{k}_{2} \bar{s}}
a^{\tau}_{\bm{k}_{2} \bar{s}}
a^{\tau}_{\bm{k}_{1} s},  \\
\hat{\text{V}}_{\text{inter}} &=
\frac{1}{N}
\sum_{\tau s_{1} s_{2}}
\sum_{\bm{k}_{1}\bm{k}_{2} }
U^{\prime}
a^{\tau \dagger}_{\bm{k}_{1} s_{1}}
a^{\bar{\tau} \dagger}_{\bar{\bm{k}}_{2} s_{2}}
a^{\tau}_{\bm{k}_{1} s_{2}}
a^{\bar{\tau}}_{\bar{\bm{k}}_{2} s_{1}}.
\end{align}
Therefore, the total Hamiltonian is obtained
\begin{equation}
\hat{\text{H}} = \hat{\text{H}}_{0} + \hat{\text{V}}_{\text{intra}} + \hat{\text{V}}_{\text{inter}},
\end{equation}
which includes the intravalley and intervalley interaction. In the following, $\hat{\text{H}}$ is solved at the mean field level.
\section{mean field approximation}

As for $\hat{\text{V}}_{\text{intra}}$, we take the mean field approximation (MFA) directly
\begin{align} \label{B1}
\hat{\text{V}}^{\text{MF}}_{\text{intra}} &=
\frac{1}{N}
\sum_{\tau s}
\sum_{\bm{k}_{1}\bm{k}_{2}}
U
\left(
\langle
a^{\tau\dagger}_{\bm{k}_{1} s}
a^{\tau}_{\bm{k}_{1} s}
\rangle
a^{\tau\dagger}_{\bm{k}_{2} \bar{s}}
a^{\tau}_{\bm{k}_{2} \bar{s}} \right. \nonumber\\
&\,\,\,\,\,\
+\left.
a^{\tau\dagger}_{\bm{k}_{1} s}
a^{\tau}_{\bm{k}_{1} s}
\langle
a^{\tau\dagger}_{\bm{k}_{2} \bar{s}}
a^{\tau}_{\bm{k}_{2} \bar{s}}
\rangle
-
\langle
a^{\tau\dagger}_{\bm{k}_{1} s}
a^{\tau}_{\bm{k}_{1} \text{s}}
\rangle
\langle
a^{\tau\dagger}_{\bm{k}_{2} \bar{s}}
a^{\tau}_{\bm{k}_{2} \bar{s}}
\rangle
\right).
\end{align}
In the mean field approximation, we neglect the second order quantum fluctuations. The third term in above equation is omitted, because it is a constant, which can not effect the following qualitative discussion of the result. MFA of $\hat{\text{V}}_{\text{intra}}$ reads
\begin{align}
\hat{\text{V}}^{\text{MF}}_{\text{intra}}
&\approx
\frac{1}{N}
\sum_{\tau s}
\sum_{\bm{k}_{1}\bm{k}_{2}}
U
\left(
\langle
n^{\tau}_{\bm{k}_{1} s}
\rangle
n^{\tau}_{\bm{k}_{2} \bar{s}}
+
n^{\tau}_{\bm{k}_{1} s}
\langle
n^{\tau}_{\bm{k}_{2} \bar{s}}
\rangle
\right) \nonumber\\
&\approx \frac{2}{N}
\sum_{\tau s}
\sum_{\bm{k}_{1}\bm{k}_{2}}
U
\langle
n^{\tau}_{\bm{k}_{2} \bar{s}}
\rangle
n^{\tau}_{\bm{k}_{1} s} \nonumber\\
&\approx
\sum_{\tau\bm{k} s}
\mathbb{U}_{\tau s}
a^{\tau \dagger}_{\bm{k} s}
a^{\tau }_{\bm{k} s}
\end{align}
where
\begin{equation} \label{MF1}
\mathbb{U}_{\tau s} =
\frac{2}{N}
\sum_{\bm{k}}
U
\langle
n^{\tau }_{\bm{k} \bar{s}}
\rangle.
\end{equation}
$n^{\tau}_{\bm{k} s} = a^{\tau \dagger}_{\bm{k} s}
a^{\tau }_{\bm{k} s}$ is the particle number operator. Here, we merely consider the zero temperature case. So, $\langle\cdots\rangle$ is the ground state average. As for $\hat{\text{V}}_{\text{inter}}$, we rewrite it in terms of the spin operators in order to extract the intervalley spin exchange interaction
\begin{equation} \label{v_inter_eq}
\hat{\text{V}}_{\text{inter}} =
-\frac{1}{\text{N}}
\sum_{\bm{k}_{1}\bm{k}_{2} }
U^{\prime}
\left(
n^{\tau}_{\bm{k}_{1} }
n^{\bar{\tau}}_{\bar{\bm{k}}_{2} }
+
4
\bm{S}_{\tau\bm{k}_{1}}
\cdot
\bm{S}_{\bar{\tau}\bar{\bm{k}}_{2}}
\right).
\end{equation}
The spin coupling term reads
\begin{equation}
\bm{S}_{\tau \bm{k}_{1}} \cdot \bm{S}_{\bar{\tau} \bar{\bm{k}}_{2}}
=
S^{\text{z}}_{\tau \bm{k}_{1}} S^{\text{z}}_{\bar{\tau} \bar{\bm{k}}_{2}} + \frac{1}{2} \left( S^{+}_{\tau \bm{k}_{1}} S^{-}_{\bar{\tau} \bar{\bm{k}}_{2}} + S^{-}_{\tau\bm{k}_{1}} \text{S}^{+}_{\bar{\tau} \bar{\bm{k}}_{2}} \right),
\end{equation}
where
$S^{ z}_{\tau \bm{k}} = \frac{1}{2} ( n^{\tau}_{\bm{k} \uparrow} - n^{\tau}_{\bm{k} \downarrow}) $
,
$S^{+}_{\tau \bm{k}} = a^{\tau \dagger}_{\bm{k} \uparrow} a^{\tau}_{\bm{k} \downarrow}$
and
$S^{-}_{\tau \bm{k}} = a^{\tau \dagger}_{\bm{k} \downarrow} a^{\tau}_{\bm{k} \uparrow}$.
It is obvious that the intervalley spin coupling is extracted. We chose the direction of $\bm{S}$ as the z-axis and apply MFA to Eq.(\ref{v_inter_eq}) obtaining
\begin{align}  \label{B6}
\hat{\text{V}}^{\text{MF}}_{\text{inter}}
&=
-\frac{1}{N}
\sum_{\bm{k}_{1}\bm{k}_{2} }
U^{\prime}
\left(
\langle
n^{\tau}_{\bm{k}_{1} }
\rangle
n^{\bar{\tau}}_{\bar{\bm{k}}_{2} }
+
n^{\tau}_{\bm{k}_{1} }
\langle
n^{\bar{\tau}}_{\bar{\bm{k}}_{2} }
\rangle
-
\langle
n^{\tau}_{\bm{k}_{1} }
\rangle
\langle
n^{\bar{\tau}}_{\bar{\bm{k}}_{2} }
\rangle \right.\nonumber\\
&\,\,\,\,\,\
\left.
+
4
\langle
\bm{S}^{z}_{\tau\bm{k}_{1}}
\rangle
\bm{S}^{z}_{\bar{\tau}\bar{\bm{k}}_{2}}
+
4
\bm{S}^{z}_{\tau\bm{k}_{1}}
\langle
\bm{S}^{z}_{\bar{\tau}\bar{\bm{k}}_{2}}
\rangle
-
4
\langle
\bm{S}^{z}_{\tau\bm{k}_{1}}
\rangle
\langle
\bm{S}^{z}_{\bar{\tau}\bar{\bm{k}}_{2}}
\rangle
\right) \nonumber\\
&\approx
- \sum_{\tau\bm{k}}
\left(
\mathbb{U}^{\prime}_{\tau}
n^{\tau}_{\bm{k}}
+
\mathbb{M}_{\tau}
S^{z}_{\tau \bm{k}}
\right)\nonumber\\
&\approx
- \sum_{\tau\bm{k}s}
\left(
\mathbb{U}^{\prime}_{\tau}
+
\frac{1}{2}s
\mathbb{M}_{\tau}
\right)
n^{\tau}_{\bm{k}s} \nonumber\\
&\approx
- \sum_{\tau\bm{k}s}
\mathbb{X}_{\tau s}
a^{\tau\dagger}_{\bm{k}s}
a^{\tau}_{\bm{k}s}
\end{align}
where
\begin{align}
\mathbb{U}^{\prime}_{\tau} &=
\frac{1}{N}
\sum_{\bm{k}}
U^{\prime}
\langle
n^{\bar{\tau}}_{\bar{\bm{k}}}
\rangle,\label{MF2_1}\\
\mathbb{M}_{\tau} &=
\frac{4}{N}
\sum_{\bm{k}}
U^{\prime}
\langle
S^{z}_{\bar{\tau} \bar{\bm{k}}} \rangle,\label{MF2_2}\\
\mathbb{X}_{\tau s} &=
\mathbb{U}^{\prime}_{\tau} + \frac{1}{2}s\mathbb{M}_{\tau}. \label{MF2_3}
\end{align}
Therefore, MFA of the interaction operator $\hat{\text{V}}$ is
\begin{align}
\hat{\text{V}}_{\text{MF}} &= \hat{\text{V}}^{\text{MF}}_{\text{intra}} + \hat{\text{V}}^{\text{MF}}_{\text{inter}} \nonumber\\
&=
\sum_{\bm{k}s}
\left(
\mathbb{F}_{\tau s}
a^{\tau\dagger}_{\bm{k}s}
a^{\tau}_{\bm{k}s}
+
\mathbb{F}_{\bar{\tau} s}
a^{\bar{\tau} \dagger}_{\bar{\bm{k}} s}
a^{\bar{\tau}}_{\bar{\bm{k}} s}
\right),
\end{align}
where the effective mean field is
\begin{equation} \label{MFtotal}
\mathbb{F}_{\tau s} = \mathbb{U}_{\tau s}
-
\mathbb{X}_{\tau s}.
\end{equation}
At the mean field level, the total hamiltonian reads
\begin{equation}
\text{H}_{\text{total}}^{\text{MF}} = \sum_{\bm{k}s}
\left(
\mathbb{E}_{c\tau s}\left(\bm{k}\right)
a^{\tau\dagger}_{\bm{k}s}
a^{\tau}_{\bm{k}s}
+
\mathbb{E}_{\text{c}\bar{\tau}s}\left(\bar{\mathbf{k}}\right)
a^{ \bar{\tau} \dagger}_{\bar{\bm{k}} s}
a^{ \bar{\tau}}_{\bar{\bm{k}} s}
\right),
\end{equation}
where the energy spectrum is
\begin{equation} \label{energy_spectrum_H}
\mathbb{E}_{c\tau s}\left(\bm{k}\right) =
E_{c \tau s}\left( \bm{k} \right)
+ \mathbb{F}_{\tau s} - \mu.
\end{equation}
It is obvious that the effective mean field $\mathbb{F}_{\tau s}$ is obtained upon the calculation of $\langle n^{\tau}_{\bm{k} s}\rangle$. It is convenient to define
\begin{equation} \label{ratio_occ}
\widetilde{n}^{\tau}_{ \text{s}} = \frac{1}{N} \sum_{\bm{k}} \langle n^{\tau}_{\bm{k} \text{s}}\rangle,
\end{equation}
which indicates the ratio of the occupation at $\tau$ valley. Eq.(\ref{energy_spectrum_H0}), Eq.(\ref{MF1}), Eq.(\ref{MF2_1}), Eq.(\ref{MF2_2}), Eq.(\ref{MF2_3}), Eq.(\ref{MFtotal}), Eq.(\ref{energy_spectrum_H}) and Eq.(\ref{ratio_occ}) constitute a set of mean field self-consistent equations.

\section{Calculations on $\widetilde{n}^{\tau}_{s}$ and $\text{E}_{\text{free}}$}
\begin{figure}[t!]
	\begin{center}
		\includegraphics[width=\columnwidth]{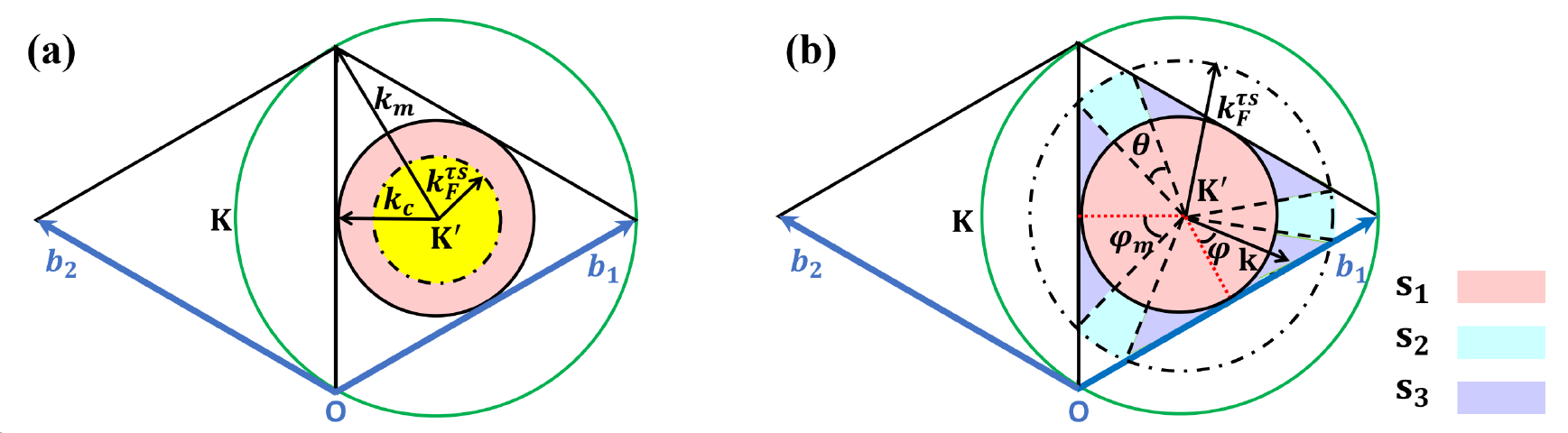}
	\end{center}
	\caption{(Color online) The BZ of ML-$\text{MoS}_{2}$. $\bm{b}_{1}$ and $\bm{b}_{2}$ are the primitive vectors of the reciprocal lattice. $\textbf{K}$ and $\textbf{K}^{\prime}$ marks two valleys. The black and green solid circle are the inscribed and circumscribed circle of the right half of BZ. The corresponding radius is $k_{c}$ and $k_{m}$ as shown in (a). The inscribed circle is filled with rose color. $k^{\tau s}_{F}$ is the Fermi radius. $\bm{k}$ is any vector in BZ. (a) $0 \le k^{\tau s}_{F} \le k_{c}$. The region within the black dash-dot circle is occupied by the electrons and filled with yellow color. (b) $k_{c} < k^{\tau s}_{F} \le k_{m}$. In this case, electrons occupy the colored region, which is composed of $s_{1}$, $s_{2}$ and $s_{3}$. $\varphi$ is the angle between $\bm{k}$ and the red dot line. $\varphi_{m}$ is the maximum of the angle. }.  \label{FigS2}
\end{figure}

In this section we calculate $\widetilde{n}^{\tau}_{s}$ and $\text{E}_{\text{free}}$. The Brillouin zone (BZ) of ML-$\text{MoS}_{2}$ is shown in Fig.~\ref{FigS2}. $\bm{b}_{1}$ and $\bm{b}_{2}$ are the two primitive vectors of BZ
\begin{equation}
\bm{b}_{1} = \left( \frac{2\pi}{\sqrt{3} a}, \frac{2\pi}{3a}\right)
,\,\,\,\,\
\bm{b}_{2} = \left( - \frac{2\pi}{\sqrt{3} a}, \frac{2\pi}{3a}\right).
\end{equation}
It is easy to obtain the area of BZ, $S_{\text{BZ}} = 8\sqrt{3}\pi^2/9a^2.$
For simplicity and compactness of the formula, we define $\mathbb{A}_{\tau s} = \lambda \tau s/2 - \mu + \mathbb{F}_{\tau s}$ and $\mathbb{B}_{\tau s} = \left(\Delta - \lambda \tau s \right)/2$. Energy spectrum is rewritten as
\begin{equation}
\mathbb{E}_{c\tau s}\left(\bm{k}\right) = \mathbb{A}_{\tau s} \pm \sqrt{(atk)^{2} + \mathbb{B}^{2}_{\tau s}}.
\end{equation}
If $\mathbb{A}_{\tau s}  \le 0$ and $|\mathbb{A}_{\tau s} | \ge | \mathbb{B}_{\tau s} |$, we are able to solve $\mathbb{E}_{\text{c}\tau\text{s}}\left( k \right) = 0$ and obtain the Fermi radius
\begin{equation}
k^{\tau s}_{F} = \frac{1}{at} \sqrt{ \mathbb{A}^{2}_{\tau s} - \mathbb{B}^{2}_{\tau s}}.
\end{equation}
As shown in Fig.~\ref{FigS2}(a), if $0 \le k^{\tau s}_{F} \le k_{c}$, electrons occupy the yellow region of BZ. If $k_{c} < k^{\tau s}_{F} \le k_{m}$, electrons occupy the colored region as shown in Fig.~\ref{FigS2}(b). $k_{c} = 2\sqrt{3}\pi/9 a$ and $k_{m} = 2k_{c}$ are the radius of the inscribed and circumscribed circle of the right half of BZ. Instead of the summation of the discrete values in Eq.(\ref{ratio_occ}), we take the value of $k$ continuously.  The definition of $\widetilde{n}^{\tau}_{s}$ is rewritten equivalently as
\begin{equation} \label{ratio_occ_conti}
\widetilde{n}^{\tau}_{s} = \frac{S^{\tau s}_{\text{occ}}}{S_{\text{BZ}}}.
\end{equation}
where $S^{\tau s}_{\text{occ}}$ is the area of the region which is occupied by the electron. When $0 \le k^{\tau s}_{F} \le k_{c}$,
\begin{equation}
\text{S}^{\tau s}_{\text{occ}} = \pi \left(k^{\tau s}_{F}\right)^{2}.
\end{equation}
Substituting into Eq.(\ref{ratio_occ_conti}), we obtain
\begin{equation}
\widetilde{n}^{\tau}_{s} = \frac{3\sqrt{3} a^{2}\left(k^{\tau s}_{ F} \right)^{2}}{8\pi}.
\end{equation}
When $\text{k}_{c} < k^{\tau s}_{F} \le k_{m}$, $S^{\tau s}_{\text{occ}}$ is divided into two parts: three triangle areas, three sectorial areas (see Fig.~\ref{FigS2}(b), regions divided by dash lines).
\begin{equation}
S^{\tau s}_{\text{occ}} = S^{\tau s}_{\text{tri}} + S^{\tau s}_{\text{sec}}.
\end{equation}
As for $S^{\tau\text{s}}_{\text{tri}}$, we have
\begin{equation}
S^{\tau s}_{\text{tri}} = 3k_{c} \sqrt{\left(k^{\tau s}_{F}\right)^{2} - k^{2}_{c}}.
\end{equation}
The sectorial area is
\begin{equation}
S^{\tau s}_{\text{sec}} = \frac{3}{2 }\left(k^{\tau s}_{F}\right)^{2}\theta,
\end{equation}
where $\theta$ is the angle of the sector. Therefore,
\begin{equation}
S^{\tau s}_{\text{occ}} = 3k_{c} \sqrt{\left(k^{\tau s}_{F}\right)^{2} - k^{2}_{c}} + \frac{3}{2 }\left(k^{\tau s}_{F}\right)^{2}\theta.
\end{equation}
Substituting into Eq.(\ref{ratio_occ_conti}), we have
\begin{equation}
\widetilde{n}^{\tau}_{s} = \frac{3a}{4\pi} \sqrt{\left(k^{\tau s}_{F}\right)^{2} - k^{2}_{c}} + \frac{\sqrt{3}}{12} \left( \frac{k^{\tau s}_{F}}{k_{c}} \right)^{2}\theta.
\end{equation}
When $k^{\tau s}_{F}>k_{m}$, $S^{\tau s}_{\text{occ}}$ is half of $S_{\text{BZ}}$. Therefore, $\widetilde{n}^{\tau}_{s} = 1/2$. Conclusively,
\begin{equation} \label{calculate_n}
\widetilde{n}^{\tau}_{s} =
\left\{
\begin{aligned}
& \frac{3\sqrt{3} a^{2}\left(k^{\tau s}_{F}\right)^{2}}{8\pi}  && 0 \le k^{\tau s}_{F} \le k_{c} \\
& \frac{3a}{4\pi} \sqrt{\left(k^{\tau s}_{F}\right)^{2} - k^{2}_{c}} + \frac{\sqrt{3}}{12} \left( \frac{k^{\tau s}_{F}}{k_{c}} \right)^{2}\theta && k_{c} < k^{\tau s}_{F} \le k_{m} \\
& \frac{1}{2} && k_{m} < k^{\tau s}_{F}
\end{aligned}
\right.
\end{equation}
where
\begin{equation}
\theta = 2 \text{arcsin}\left( \frac{\pi}{3ak^{\tau s}_{F}} - \frac{1}{2k^{\tau s}_{ F}} \sqrt{ \left( k^{\tau s}_{F} \right)^{2} - k^{2}_{c}} \right).
\end{equation}

As for the electron which fills the conduction band, its contribution to the free energy is defined by the integration
\begin{equation} \label{FEPrime}
\text{E}^{\tau s}_{\text{free}} = \int_{S^{\tau s}_{\text{occ}}} d\bm{k}
\left(
\mathbb{A}_{\tau s} + \sqrt{\left( atk\right)^{2} + \mathbb{B}^{2}_{\tau s}}
\right).
\end{equation}
Hence, the total free energy reads
\begin{equation} \label{FE}
\text{E}_{\text{free}} = \sum_{\tau s} \text{E}^{\tau s}_{\text{free}}.
\end{equation}
In Eq. (\ref{FEPrime}), we neglect the wave vector density.
When $0 \le k^{\tau s}_{F} \le k_{c}$, the region occupied by the electrons in BZ is a circular region. $\text{E}^{\tau s}_{\text{free}}$ is calculated directly.

\begin{align} \label{FE1}
\text{E}^{\tau s}_{\text{free}}
&= -\int^{\mathbb{E}_{c\tau s} \left(k^{\tau s}_{F}\right)}_{\mathbb{E}_{c\tau s}(0)}
\pi k^{2}
d\mathbb{E} \nonumber\\
&= -\int^{\mathbb{E}_{c\tau s} \left(k^{\tau s}_{F}\right)}_{\mathbb{E}_{c\tau s}(0)}
\frac{\pi}{(at)^{2}}
\left[
\left(
\mathbb{E} - \mathbb{A}_{\tau s}
\right)^{2}
-
\mathbb{B}^{2}_{\tau s}
\right]
d\mathbb{E} \nonumber\\
&= \frac{\pi}{3\left( at\right)^{2}}
\left(
\mathbb{A}_{\tau s}^{3}
-2\mathbb{B}^{3}_{\tau s}
-3\mathbb{B}^{2}_{\tau s} \mathbb{A}_{\tau s}
\right).
\end{align}
In above derivation, we use $\mathbb{E}\left(k^{\tau s}_{F}\right) = 0$ and $\mathbb{E}(0) = \mathbb{A}_{\tau s} + \mathbb{B}_{\tau s}$.
As for the case $k_{c} < \text{k}^{\tau s}_{F} \le k_{m}$, $E^{\tau s}_{free}$ is composed of three parts, which are corresponding to the integration over the region $s_{1}$, $s_{2}$ and $s_{3}$ as shown in Fig.~\ref{FigS2}(b),
\begin{equation} \label{FE2}
\text{E}^{\tau s}_{\text{free}} =
\text{E}^{\tau s}_{1} +
\text{E}^{\tau s}_{2} +
\text{E}^{\tau s}_{3}.
\end{equation}
We calculate the integration individually.

\emph{(1)} The integration in region $s_{1}$ is
\begin{align} \label{FE2_1}
\text{E}^{\tau s}_{1}
&= -\int^{\mathbb{E}_{c\tau s} \left(k_{c}\right)}_{\mathbb{E}_{c\tau s}(0)}
\pi k^{2}
d\mathbb{E}
-
\int^{\mathbb{E}_{c\tau s} \left(k^{\tau s}_{F}\right)}_{\mathbb{E}_{c\tau s} \left(k_{c}\right)}
\pi k^{2}_{c}
d\mathbb{E} \nonumber\\
&= -\int^{\mathbb{E}_{c\tau s} \left(k_{c}\right)}_{\mathbb{E}_{c\tau s}(0)}  \frac{\pi}{(at)^{2}}
\left[
\left(
\mathbb{E} - \mathbb{A}_{\tau s}
\right)^{2}
-
\mathbb{B}^{2}_{\tau s}
\right]
d\mathbb{E} \nonumber\\
&\,\,\,\,\,\
+
\pi k^{2}_{c} \mathbb{E}_{c\tau s}(k_{c})
\nonumber\\
&= \frac{\pi}{3(at)^{2}}
\left(
3 \mathbb{B}^{2}_{\tau s} \mathbb{C}_{\tau s} -
\mathbb{C}^{3}_{\tau s} - 2 \mathbb{B}^{3}_{\tau s}
\right) \nonumber\\
&\,\,\,\,\,\
+
\pi k^{2}_{c} \mathbb{E}_{c\tau s}(k_{c}),
\end{align}
where
\begin{equation}
\mathbb{C}_{\tau s} = \sqrt{ \left( at k_{c} \right)^{2} + \mathbb{B}^{2}_{\tau s}}.
\end{equation}

\emph{(2)} The integration in region $s_{2}$ is
\begin{align}
\text{E}^{\tau s}_{2} &=
\frac{3\theta}{2\pi}
\left(
-\int^{\mathbb{E}_{c\tau s}
\left( k^{\tau s}_{F} \right)}_{\mathbb{E}_{c \tau s}(0)}
\pi k^{2}
d\mathbb{E}
-
\text{E}^{\tau s}_{1}
\right) \nonumber\\
&=\frac{\theta}{2(at)^{2}}
\left(
\mathbb{A}^{3}_{\tau s}
-
2 \mathbb{B}^{3}_{\tau s}
-
3 \mathbb{B}^{2}_{\tau s} \mathbb{A}_{\tau s}
\right)
-
\frac{3\theta}{2\pi}
\text{E}^{\tau s}_{1}.
\end{align}

\emph{(3)} The integration in region $s_{3}$ is complicated. We have
\begin{align}
\text{E}^{\tau s}_{3} &=
6\int_{s_{3}}
\mathbb{A}_{\tau s}
+
\sqrt{(atk)^{2} + \mathbb{B}^{2}_{\tau s}}
dk_{x} dk_{y} \nonumber \\
&=
6\int^{\varphi_{m}}_{0}
d\varphi
\int^{k_{c}/\text{cos}(\varphi)}_{k_{c}}
k
\left(
\mathbb{A}_{\tau s}
+
\sqrt{(atk)^{2} + \mathbb{B}^{2}_{\tau s}}
\right)
dk \nonumber \\
&=6\int^{\varphi_{m}}_{0}
d\varphi
\left\{
\frac{1}{2} \mathbb{A}_{\tau s} k^{2}_{c}
\left(
\frac{1}{\text{cos}^{2}(\varphi)} - 1
\right)  \right.\nonumber \\
&\,\,\,\,\,\
+
\frac{1}{3(at)^{2}}
\left[
\left(
\frac{atk_{c}}{\text{cos}\left( \varphi \right)}
\right)^{2}
+
\mathbb{B}^{2}_{\tau s}
\right]^{\frac{3}{2}} \nonumber\\
&\,\,\,\,\,\
-
\left.
\frac{1}{3(at)^{2}}
\left[
\left(
atk_{c}
\right)^{2}
+
\mathbb{B}^{2}_{\tau s}
\right]^{\frac{3}{2}}
\right\}\nonumber \\
&=3
\mathbb{A}_{\tau s} k^{2}_{c}
\left(
\text{tan}\left( \varphi_{m} \right) - \varphi_{m}
\right) -\frac{2}{\left(at\right)^{2}} \varphi_{m} \mathbb{C}^{3}_{\tau s} \nonumber \\
&\,\,\,\,\,\,
+ 6 \mathbb{D}_{\tau s},
\end{align}
where
$\varphi_{m} = \text{arccos}\left( k_{c}/k^{\tau s}_{F} \right)$. $\mathbb{D}_{\tau s}$ denotes an integration
\begin{equation}
\mathbb{D}_{\tau s} =
\frac{1}{3\left( at \right)^{2}}
\int_{0}^{\varphi_{m}} d\varphi
\left[
\left(
\frac{atk_{c}}{\text{cos}\left( \varphi \right)}
\right)^{2}
+
\mathbb{B}^{2}_{\tau s}
\right]^{\frac{3}{2}}.
\end{equation}
It is hard for $\mathbb{D}_{\tau s} $ to obtain an analytical formula. Therefore, $\mathbb{D}_{\tau s}$ is calculated numerically.
When $k_{c} < k^{ \tau s }_{F} \le k_{m}$,
\begin{align}
\text{E}^{\tau s}_{\text{free}} &=
\frac{\pi}{3(at)^{2}}
\left(
3 \mathbb{B}^{2}_{\tau s} \mathbb{C}_{\tau s} -
\mathbb{C}^{3}_{\tau s} - 2 \mathbb{B}^{3}_{\tau s}
\right)
+
\pi k^{2}_{c} \mathbb{E}_{\text{c}\tau s}(k_{c}) \nonumber\\
&\,\,\,\,\,\
+
\frac{\theta}{2(at)^{2}}
\left(
\mathbb{A}^{3}_{\tau s}
-
2 \mathbb{B}^{3}_{\tau s}
-
3 \mathbb{B}^{2}_{\tau s} \mathbb{A}_{\tau s}
\right)
-
\frac{3\theta}{2\pi}
\text{E}^{\tau s}_{1} \nonumber\\
&\,\,\,\,\,\
+
3
\mathbb{A}_{\tau s} k^{2}_{c}
\left(
\text{tan}\left( \varphi_{m} \right) - \varphi_{m}
\right) -\frac{2}{\left(at\right)^{2}} \varphi_{m} \mathbb{C}^{3}_{\tau s} \nonumber\\
&\,\,\,\,\,\
+
6 \mathbb{D}_{\tau s}.
\end{align}
As for $0 \le k^{\tau s}_{F} \le k_{c}$, the total free energy is obtained by substituting Eq. (\ref{FE1}) into Eq. (\ref{FE}). For $k_{c} < k^{ \tau s }_{F} \le k_{m}$, the total free energy is calculated by substituting Eq. (\ref{FE2}) into Eq. (\ref{FE}).


%


\end{document}